\documentclass[fleqn,usenatbib]{mnras}
\usepackage{newtxtext,newtxmath}
\usepackage[T1]{fontenc}
\usepackage{ae,aecompl}
\usepackage{xcolor} 
\usepackage{multicol}        
\usepackage{subfigure}
\usepackage{bm}		
\usepackage{graphicx}	
\usepackage{amsmath}	
\usepackage{footmisc}   
\defcitealias{duguid_tidal_2019}{Paper I} 
\title[Tides and convection.]{Convective turbulent viscosity acting on equilibrium tidal flows: new frequency scaling of the effective viscosity}
\author[C. D. Duguid et al.]{
	Craig. D. Duguid,$^{1}$\thanks{E-mail: sccd@leeds.ac.uk}
	Adrian. J. Barker,$^{2}$
	and C. A. Jones $^{2}$
	\\
	$^{1}$EPSRC Centre for Doctoral Training in Fluid Dynamics, University of Leeds, Leeds LS2 9JT, UK\\
	$^{2}$School of Mathematics, University of Leeds, Leeds LS2 9JT, UK\\
}
\date{Accepted XXX. Received YYY; in original form ZZZ}
\pubyear{2020}

\begin{document}
	\label{firstpage}
	\pagerange{\pageref{firstpage}--\pageref{lastpage}}
	\maketitle
	\begin{abstract}
		Turbulent convection is thought to act as an effective viscosity ($\nu_E$) in damping tidal flows in stars and giant planets. However, the efficiency of this mechanism has long been debated, particularly in the regime of fast tides, when the tidal frequency ($\omega$) exceeds the turnover frequency of the dominant convective eddies ($\omega_c$). We present the results of hydrodynamical simulations to study the interaction between tidal flows and convection in a small patch of a convection zone. These simulations build upon our prior work by simulating more turbulent convection in larger horizontal boxes, and here we explore a wider range of parameters. We obtain several new results: 1) $\nu_E$ is frequency-dependent, scaling as $\omega^{-0.5}$ when $\omega/\omega_c \lesssim 1$, and appears to attain its maximum constant value only for very small frequencies ($\omega/\omega_c \lesssim 10^{-2}$). This frequency-reduction for low frequency tidal forcing has never been observed previously. 2) The frequency-dependence of $\nu_E$ appears to follow the same scaling as the frequency spectrum of the energy (or Reynolds stress) for low and intermediate frequencies. 3) For high frequencies ($\omega/\omega_c\gtrsim 1-5$), $\nu_E\propto \omega^{-2}$. 4) The energetically-dominant convective modes always appear to contribute the most to $\nu_E$, rather than the resonant eddies in a Kolmogorov cascade. These results have important implications for tidal dissipation in convection zones of stars and planets, and indicate that the classical tidal theory of the equilibrium tide in stars and giant planets should be revisited. We briefly touch upon the implications for planetary orbital decay around evolving stars. 
	\end{abstract}
	
	

	\section{Introduction}	
	Understanding tidal interactions is important because they drive spin and orbital evolution in a wide range of astrophysical systems. In many such systems the influence of tides has been inferred, such as from the distribution of eccentricities \citep{Meibom2005,Geller2012,VanEylen2016,Triaud2017,Nine2020} and spin periods of close binaries \citep{Meibom2006,Lurie2017}.
	In other cases the effects of tides have been directly observed, such as the tidally-excited oscillations of heartbeat stars \citep{Welsh2011,Zimmerman2017,Guo2020}, the tidally-driven inspiral of WASP-12b \citep{Maciejewski2016,Maciejewski2018,Patra2017,yee_orbit_2019}, or for various features of moons within the Solar system (\citealt{deeg_interiors_2017} and references therein).
	With such a diversity of applications, there is a growing interest in improving tidal theory.
	
	There are currently many aspects of tidal interactions that are active areas of research, and in particular, mechanisms relating to the dissipation of tidal energy. For example, tidal excitation of internal gravity waves \citep[e.g.][]{GoodmanDickson1998,OgilvieLin2007,BO2010,B2011,Weinberg2012,Essick2016} can be important in stellar radiation zones, and this mechanism may be responsible for the observed orbital decay of WASP-12 b \citep{Maciejewski2016,Patra2017,Chernov2017,Weinberg2017,Bailey2019,yee_orbit_2019}. The excitation of inertial waves is being studied in the convective envelopes of rotating stars or giant planets \citep[e.g.][]{Wu2005,OgilvieLin2007,GoodmanLackner2009,PapIv2010,FBBO2014,B2016}, and this mechanism may be important for tidal circularisation and spin synchronisation. In giant planets, the role of stably-stratified (or semi-convective) layers is also being explored \citep{fuller_resonance_2016,andre_layered_2017,Andre2019,pontin_wave_2020} with possible application to the orbital migration of the moons of Jupiter and Saturn \citep[e.g.][]{Lainey2009,Lainey2012,Lainey2017,Lainey2020}.
	
	In this work we focus on tidal dissipation resulting from the interaction of large-scale (non-wavelike) equilibrium tides and convection inside stars or giant planets. This is a classical tidal mechanism that is commonly believed to be important in stars (or giant planets) with convective envelopes \citep{zahn_les_1966,Zahn1989,Zahn2008}. The interaction between the tide and convection is thought to act like an effective viscosity $\nu_E$ (which is much larger than the microscopic viscosity) in damping the large-scale tidal flow. However, the efficiency of this mechanism is expected to be reduced when the tidal shear frequency $\omega$ exceeds that of the relevant convective frequency $\omega_c$, but the power law of this reduction has long been a matter of debate \citep[e.g.][]{GoodmanOh1997}, and this issue is still often considered as ``the Achilles' heel of tidal theory" \citep{Zahn2008}. Using a refinement of mixing-length theory, \cite{zahn_les_1966} used phenomenological arguments to predict that $\nu_E \sim (\omega/\omega_c)^{-1}$ when $\omega/\omega_c\gg 1$, while, by applying similar ideas to a Kolmogorov turbulent cascade, \cite{goldreich_turbulent_1977} predicted $\nu_E \sim(\omega/\omega_c)^{-2}$ when $\omega/\omega_c\gg 1$. It is essential to determine which of these (if either) are correct because the corresponding timescales of tidal evolution can differ by many orders of magnitude between these two prescriptions \citep[e.g.][]{duguid_tidal_2019}.
	
	Modern numerical techniques and computational power allow the dissipation of the equilibrium tide to be explored through numerical experiments. Although this has not yet completely settled the issue, the evidence in favour of the $-2$ power law of \cite{goldreich_turbulent_1977} for $\omega/\omega_c\gg 1$ has been mounting in recent years using a variety of models. The work of \cite{goldman_effective_2008} followed the ideas of \cite{goldreich_turbulent_1977} by applying an idealised turbulence model to obtain an analytical expression in agreement with the $-2$ power law. An asymptotic theory valid in the limit of high frequency tidal forcing has been developed \citep{ogilvie_interaction_2012,duguid_tidal_2019}, which robustly predicts a $-2$ scaling (and its predictions have been verified using laminar simulations; \citealt{braviner_stellar_2015,duguid_tidal_2019}).
	Various simulations have been performed which support the $-2$ scaling, such as homogeneous convection in a triply-periodic Cartesian domain \citep{ogilvie_interaction_2012}, the convection analog of ABC flow in a Cartesian domain \citep{braviner_stellar_2015}, Rayleigh-B\'{e}nard convection in a Cartesian domain with horizontally-periodic and stress-free boundary conditions in the vertical direction \citep{duguid_tidal_2019}, and convection in a full sphere which is homogeneously heated \citep{vidal_turbulent_2020}. With such a diversity of models one would be forgiven for assuming the matter is settled. However, \cite{penev_direct_2009} observed a $-1$ law as proposed by \cite{zahn_les_1966} for a limited range of tidal frequencies spanning $0.5 \lesssim \omega/\omega_c \lesssim 4$. Their model is unique in that they allowed for multiple density scale heights within the domain. More recently, there is evidence that both the $-1$ and $-2$ power laws may coexist in (Boussinesq) spherical convection \citep{vidal_turbulent_2020}. However, in both of these latter studies, the $-1$ power law was observed only for intermediate frequencies $\omega\sim \omega_c$, and not for $\omega/\omega_c\gg 1$ as originally proposed by \cite{zahn_les_1966}.
	
	The purpose of this study is to build upon \cite{duguid_tidal_2019} (hereafter \citetalias{duguid_tidal_2019}), which explored the interaction between Rayleigh-B\'{e}nard convection and an oscillatory tidal-like flow in Cartesian domains, by exploring larger domain sizes and by performing a much wider parameter survey. The former is important because our previous results suggested that the energetically-dominant modes of the convection (the large-scale modes) contributed the most to the effective viscosity, and we desire to obtain results that are independent of the size of our Cartesian domain. Our wider parameter survey will also enable us to explore the effective viscosity not just for the highest forcing frequencies, but over a wide range of astrophysically-relevant frequencies.
	
	Despite the results of \citetalias{duguid_tidal_2019} agreeing with the power law proposed by \cite{goldreich_turbulent_1977} for high frequencies, our observation that the largest scales dominated the effective viscosity differs from the mechanism proposed in their theory. Indeed, in \citetalias{duguid_tidal_2019}, we found that the largest scales were comparable with the domain size, and thus an investigation into the impact of any constraints of adopting such a limited domain size is important. As well as investigating larger domains, and guided by \cite{penev_direct_2009} and \cite{vidal_turbulent_2020} we explore further the turbulent statistics of the convection, both with and without the tidal shear. This will enable us to determine if the nature of the turbulence, and its statistical properties, is key to understanding the frequency-dependence of the effective viscosity. If so, this would provide an important advance in our understanding of this mechanism.

	This paper is structured as follows. In section \ref{section_methods} we describe the model used in this investigation. This includes the governing equations and mathematical descriptions of the diagnostics that we have employed. In section \ref{section_results} we analyse the results of our extensive parameter survey. We begin with an exploration of unsheared convection, which lays the groundwork for the analysis of simulations with an imposed tidal-like shear flow. Analysis of the key quantities is performed throughout. In section \ref{section_discussion} we discuss the importance of the results, highlighting our most important findings and we then consider some astrophysical implications in section \ref{Implications}. We then conclude in section \ref{section_conclusions}.

	\section{Model setup}\label{section_methods}
	We follow the approach used in our previous work, which we shall briefly summarise here (see \citetalias{duguid_tidal_2019} for further details). We consider a small
	Cartesian patch of the convective envelope of a star with a large-scale non-wavelike tidal (shear) flow.
	In order to model the convection we adopt the Boussinesq approximation \citep{spiegel_boussinesq_1960} and the classical Rayleigh-B\'{e}nard setup \citep{chandrasekhar_hydrodynamic_2013}.
	Our Cartesian coordinates and domain are defined such that $x\in[0,L_x d]$, $y\in[0,L_yd]$ and $z\in[0,d]$, where $z$ is the local radial direction and $d$ is the layer depth (which strictly must be small relative to the local pressure scale height) and $x$ and $y$ represent the two horizontal directions. The boundary conditions are shearing-periodic in $x$ and periodic in $y$
	with stress-free, impermeable, and fixed temperature walls on the top ($z=d$) and bottom ($z=0$).
	
	Tidal deformations of stars are typically small and we are mainly interested in the regime of linear tides in this work\footnote{We neglect nonlinear tidal effects such as the elliptical instability in convection zones \citep{BL2013,B2016}, which might be important for tides in the shortest-period hot Jupiters.}. As such, we can study the effect of each component of the full tidal potential on the fluid separately. Consequently, we choose an oscillatory shear flow which is linear in the local coordinates (\citealt{ogilvie_interaction_2012}), and represent this flow as a `background flow' of the form 
	\begin{equation}\label{maths_shear}
		\boldsymbol{u}_0 = \frac{a_0 \omega}{d} x \cos(\omega t) \boldsymbol{e}_y \,,
	\end{equation}
	where $a_0 \omega/d$ is the amplitude of the tidal shear. The amplitude of the tidal displacement is $a_0$, and the tidal frequency is $\omega$.
	
	We solve the momentum equation in terms of the perturbed velocity ($\boldsymbol{u}$) about our tidal-like background flow ($\boldsymbol{u}_0$), where the total velocity is $\boldsymbol{u} + \boldsymbol{u}_0$. This is coupled with the thermal energy equation, which we solve for a temperature perturbation, $\theta$, about a linear background temperature profile (the conduction state). This temperature perturbation is defined as $\theta = \alpha g T$, where $\alpha$ is the thermal expansion coefficient, $g$ is the gravitational acceleration and $T$ is the ``actual" temperature perturbation. We non-dimensionalise the equations using the thermal time-scale, $d^2/\kappa$, where $\kappa$ is the (constant) thermal diffusivity, as our unit of time and $d$ as our unit of length. We will also later re-interpret our results in terms of the convective (``free-fall") time.	Thus the non-dimensional form of the governing equations under the Boussinesq approximation are
	\begin{subequations}
		\begin{align}
			&	\dfrac{\partial \boldsymbol{u}}{\partial t} + \boldsymbol{u}\cdot\nabla\boldsymbol{u} + \boldsymbol{u}_0 \cdot \nabla \boldsymbol{u} + \boldsymbol{u} \cdot \nabla \boldsymbol{u}_0 = -\nabla P + {Ra}\,{Pr} \theta \boldsymbol{e}_z + {Pr}\, \nabla^2 \boldsymbol{u} \,, \\
			&	\dfrac{\partial \theta}{\partial t} + \boldsymbol{u} \cdot \nabla \theta + \boldsymbol{u}_0 \cdot \nabla \theta = u_z + \nabla^2 \theta \,, \\
			&	\nabla \cdot \boldsymbol{u}=0 \,, \\
			& u_z = \theta = \partial_z u_x = \partial_z u_y = 0 \quad \text{ on } \quad z = 0 \text{ and } z=1 \,,
		\end{align}
	\end{subequations}
	where $P$ is the pressure and we have neglected introducing new symbols for dimensionless quantities for simplicity. The equations contain two non-dimensional parameters, the Rayleigh number ($Ra$, which measures the strength of convective driving to diffusive processes) and the Prandtl ($Pr$) number (the ratio of viscous to thermal diffusion), which were defined in \citetalias{duguid_tidal_2019}. In the interests of limiting our parameter survey we fix the Prandtl number in this work to $Pr = 1$. We also define the scaled Rayleigh number as $R={Ra} / {Ra}_c$ where ${Ra}_c$ is the critical Rayleigh number for the onset of (un-sheared)\footnote{We have found no evidence that the shear strongly modifies the onset Rayleigh number, as noted in \citetalias{duguid_tidal_2019}.} convection given our geometry and boundary conditions (\citealt{chandrasekhar_hydrodynamic_2013}). The equations are solved numerically using the Cartesian pseudo-spectral code Snoopy \citep{Lesur2005,lesur_angular_2010}, which uses time-dependent wavevectors to deal with the linear spatial dependence of $\boldsymbol{u}_0$. For more details see \citetalias{duguid_tidal_2019}.

	\subsection{Quantities of interest}\label{section_QoI}
	We seek to explore the frequency dependence of the effective (also known as eddy or turbulent) viscosity which arises as a result of the interaction between the oscillatory background tidal (shear) flow and the convective motion. This effective viscosity can be related to tidal quality factors (\citealt{ogilvie_tidal_2014}), and so it is relevant for determining the tidal evolution of astrophysical bodies. We evaluate this by defining the effective viscosity as	(\citealt{GoodmanOh1997,ogilvie_interaction_2012, braviner_stellar_2015}; \citetalias{duguid_tidal_2019})

	\begin{equation}\label{maths_eff_visc}
		\nu_E (\omega) = \frac{-2}{a_0 \omega (T - T_0)} \int_{T_0}^{T} R_{xy}(t) \cos(\omega t) \, \text{d}t \,,
	\end{equation}
	we integrate over a suitable period of time $T - T_0$ covering many tidal periods, and $R_{xy}$ is the Reynolds stress. The Reynolds stress determines the energy transfer rate between the shear and the convection, which can in principle operate in either direction (see \citealt{ogilvie_interaction_2012} and \citetalias{duguid_tidal_2019}), transferring energy from (to) the convection to (from) the shear, and is defined as
	\begin{equation}
		R_{xy}(t) = \frac{1}{V} \int_V u_x u_y \, \text{d} V\,,
	\end{equation}
	where $V = L_x L_y d$ (and $d=1$ with our non-dimensionalisation) is the volume of our domain.
	
	We define the convective frequency as $\omega_c = u_z^{\text{rms}} /d$, where $u_z^{\text{rms}}$ represents the time-averaged root-mean-square vertical velocity. When the tidal frequency is much smaller than the convective frequency, $\omega \ll \omega_c$, the convection `feels' the tidal shear as an effectively constant background shear flow due to the large differences in flow timescales. In such a situation the effective viscosity could be expected to scale like the eddy viscosity of convection as predicted by mixing-length theory (MLT). In the MLT formulation \citep[e.g.][]{BohmVitense1958,zahn_les_1966,Zahn1989} the effective viscosity is calculated using $\nu_E^{\text{mlt}} \propto u^{\text{mlt}} l^{\text{mlt}}$, where $u^{\text{mlt}}$ is the convective velocity (which is typically some relevant statistic of the vertical/radial component of velocity), $l^{\text{mlt}}$ is the mixing length (in stellar convection this is typically a multiple of the pressure scale height) and there exists a constant of proportionality. In our simulations, we define this constant of proportionality so that\footnote{Note that this strictly differs from the usual  mixing length ``$\alpha$" parameter, since it combines the usual parameter with the coefficient involved in converting $u^{\text{mlt}} l^{\text{mlt}}$ to a viscosity -- which is commonly assumed to be 1/3 without rigorous justification.\label{footnote_alpha}}
	\begin{equation}\label{maths_mlt}
		\alpha = \frac{\nu_E}{u_z^{\text{rms}} d} \,.
	\end{equation}

	As with our previous work (\citetalias{duguid_tidal_2019}) we will find it helpful to evaluate the 1D horizontal wavenumber (spatial) spectrum. To obtain this we first
	Fourier transform the three velocity components in the
	two horizontal directions to obtain the discrete version of
	($j=x, y$ or $z$)
	\begin{equation}
		\label{horizontal2D_wavenumber_spectrum}
		\hat{u}_j(k_x,k_y,z,t) = \int_{-\infty}^{\infty} \int_{-\infty}^{\infty}  u_j(x,y,z,t) e^{i(k_x x+k_y y)} \, \text{d}x \, \text{d}y .
	\end{equation}
	The 1D horizontal energy spectrum is then defined
	by writing $k_x=k_\perp \cos \theta$, 
	$k_y=k_\perp \sin \theta$, time-averaging
	and vertically integrating, so that
	\begin{equation}
		\label{exact_1D_spectrum}
		\hat{E}(k_{\perp}) = \lim_{T \to \infty} \frac{1}{2T}  \int_{0}^T \int_0^1  \int_{0}^{2\pi} ( \hat{u}_x \hat{u}_x^{\ast}  + \hat{u}_y \hat{u}_y^{\ast} + \hat{u}_z \hat{u}_z^{\ast}) \, k_\perp \,\text{d}\theta  \, \text{d}z\, \text{d}t.
	\end{equation}
	Here $\hat{\bullet}$ defines Fourier transformed quantities, $\bullet^{\ast}$ defines complex conjugates.

	In our simulations, $k_x$ and $k_y$ take on the discrete values
	\begin{equation}
		k_x = \frac{2\pi n_x}{L_x}, \quad k_y = \frac{2\pi n_y}{L_y} ,\label{discrete_wavenumbers}
	\end{equation}
	where $n_x$ and $n_y$ are integers (smaller than or equal to $N_x/2=N_y/2$), so we approximate
	eq.~\ref{exact_1D_spectrum} by considering rings with fixed width in wavenumber space. In simulations with various $L_x=L_y$ for each $R$ we select the number of $k_{\perp}$ values $N_{\perp} = N_x^{\text{largest}}$ where $N_x^{\text{largest}}$ is the number of $k_x$ values in the largest domain with length $L_x^{\text{largest}}$. We also pick the maximum $k_x$ value in the largest domain 
	as the largest value to evaluate for $k_{\perp}$. We then define the set of $k_{\perp}$ values to be 
	\begin{equation}
		k_{\perp} = \left\{ \frac{2 \pi n_{\perp}}{L_x^{\text{largest}}} : n_{\perp} \in \mathbb{N}_{0} < N_{\perp} 
		\right\}    \,,
	\end{equation}
	so that each ring has width $2\pi/L_x^{\text{largest}}$. 
	For each ring we identify the set $K$ of
	integer pairs $(n_x,n_y)$ such that 
	$(k_x, k_y)$ 
	lies inside the $k_\perp$ ring,
	\begin{equation}
	\label{set_integer_pairs}
	K = \left\{ (n_x,n_y): 
	\frac{2 \pi n_\perp}{L_x^{\text{largest}}} \leq 
	2 \pi \sqrt{\frac{n_x^2}{L_x^2} + \frac{n_y^2}{L_y^2}}
	<  \frac{2 \pi (n_\perp + 1)}{L_x^{\text{largest}}}	 \right\}.
	\end{equation}
	We note that some of the $k_{\perp}$ bins contain no $k_x\,, k_y$ values, and in such cases we remove this bin and interpolate between the adjacent bins.

	Our numerical approximation to the 1D horizontal 
	energy spectrum is then
	\begin{equation}
		\label{numerical_1D_spectrum}
		\hat{E}(k_{\perp}) = \frac{1}{2(T-T_0)}  \sum_{(n_x,n_y) \in K} \int_{T_0}^T \int_0^1  \hat{u}_x \hat{u}_x^{\ast}  + \hat{u}_y \hat{u}_y^{\ast} + \hat{u}_z \hat{u}_z^{\ast}  \, \text{d}z\, \text{d}t. 
	\end{equation}
	Note that every wavenumber pair $(k_x,k_y)$ falls in exactly one ring, so the sum of
	$\hat{E}(k_{\perp})$ over all $k_\perp$ rings exactly equals the sum over all horizontal wavenumber pairs. Our algorithm therefore partitions all of the temporally-averaged and vertically-integrated energy into bins which correspond approximately to a horizontal wavelength $2\pi/k_\perp$.  This allows us to examine the energy contained in the various horizontal length-scales.

	Previous numerical work (\citealt{penev_direct_2009}; \citetalias{duguid_tidal_2019}; \citealt{vidal_turbulent_2020}) has suggested that the frequency (temporal) spectrum of the kinetic energy, and/or Reynolds stress, may be important for determining the frequency dependence of the effective viscosity. In particular, it appears that for a certain ``intermediate" range of frequencies (meaning for an interval around $\omega\sim \omega_c$), the Reynolds stress frequency spectrum may have the same frequency dependence as the effective viscosity \citep{vidal_turbulent_2020}, though this has not yet been demonstrated for low frequencies, and the spectrum may depend on the nature of the convection \citep[e.g.][]{penev_direct_2009}. Therefore, we evaluate the frequency spectrum, which is a commonly used diagnostic in turbulent convection (e.g \citealt{ashkenazi_spectra_1999,kumar_energy_2014,kumar_applicability_2018}), by computing
	\begin{equation}\label{maths_freq_spectrum}
		\tilde{\Gamma}(\tilde{\omega}) =   \int_{-\infty}^{\infty} H(t) \, \langle \Gamma \rangle(t)\, e^{i \tilde{\omega} t} \, \text{d} t \quad \text{with} \quad \Gamma = (E , R_{xy} )
	\end{equation}
	where $\langle \Gamma \rangle$ is the volume-averaged kinetic energy $E$ or Reynolds stress $R_{xy}$, $H$ is the Hann window function \citep{oppenheim_discrete-time_2010} which we have applied in order to reduce spectral leakage and $\tilde{\omega}$ is the angular frequency. We present the frequency spectra with application of a $20$-point moving average in order to clean up the signal. We will later plot these spectra by scaling $\tilde{\omega}$ by the convective frequency $\omega_c$.
	
	Despite our previous work (\citetalias{duguid_tidal_2019}; \citealt{vidal_turbulent_2020}), and that of \citep{ogilvie_interaction_2012,braviner_stellar_2015} finding agreement with \cite{goldreich_turbulent_1977} that $\nu_E\propto(\omega_c/\omega)^2$ for high-frequency tidal forcing, i.e. $\omega\gg \omega_c$, the 
	mechanism they proposed has not been explored in detail. In particular, 
	\cite{goldreich_turbulent_1977} proposed that with a short tidal forcing timescale $\tau_T$ the resonant eddies
	would have a small length scale $\lambda$ and small typical velocity $u_\lambda$ corresponding to the values expected
	in a Kolmogorov cascade, $\lambda / l^{\text{mlt}} \sim (\tau_T/\tau_{\text{conv}})^{3/2}$ and 
	$u_{\lambda} / u^{\text{mlt}} \sim (\tau_T/\tau_{\text{conv}})^{1/2}$ giving an
	effective eddy viscosity $\nu_E \sim \lambda u_{\lambda} \sim (\tau_T/\tau_{\text{conv}})^{2} \propto 
	(\omega_c/\omega)^2	$. 
	However, \citetalias{duguid_tidal_2019} hinted that this argument may not be correct, since the large scale energetically-dominant convective modes appear to contribute the most to the effective viscosity, and the contributions appeared to fall off rapidly with increasing wavenumber. One shortcoming of our previous analysis was that the convection was intentionally simulated in a small horizontal domain (to enable a more straightforward comparison with asymptotic theory), but this artificially constrained the turbulent state, as the most energetically-dominant modes were always at the box scale in these simulations. In this paper, we revisit this issue with simulations in wider horizontal domains that are ``more turbulent", and present an analysis of the time-averaged and vertically-integrated wavenumber (spatial) spectrum of the kinetic energy, $\hat{E}(n_x,n_y)$, and Reynolds stress, $\hat{R}_{xy}(n_x,n_y)$, where $n_i$ are the integer wavenumbers $n_i = k_i L_i / \pi$, and $i=x$ or $y$. With this we are able to evaluate contributions to the effective viscosity from each wavenumber in the flow, enabling us to directly test the mechanism proposed by \cite{goldreich_turbulent_1977}.

	\subsection{Parameter survey}
	
	In this work we explore the behaviour of four of the key parameters in the problem. Our main focus is the frequency dependence of the effective viscosity $\nu_E(\omega)$.  Our new simulations build upon \citetalias{duguid_tidal_2019} by simulating wider horizontal domains ($L_x,L_y>2$), leading to ``more turbulent" convection for a given $R$, and by exploring further the low frequency regime, $\omega< \omega_c$. The parameters of our simulations are summarised in Table~\ref{table_allsims}. The data for $L_x = L_y = 2$ is the same as that presented in \citetalias{duguid_tidal_2019}, which brings the total number of simulations performed in this study to be in excess of 700. Details of some of the simulations are
	given in Table~\ref{table_allsims}. The strength of the convection is varied by varying $R$.	Due to the demanding nature of these simulations, which for convergence of $\nu_E$ are required to be integrated for multiple tidal periods (in some cases this means hundreds of diffusion times), we are limited to
	values of $R \leq 1000$, which is much smaller than the values expected in stars\footnote{The convection zones of Sun-like stars are expected to have $Ra\in [10^{21},10^{24}]$ and $Pr \in [10^{-7},10^{-3}]$ \citep[e.g.][]{hanasoge2016}. }. We therefore hope to find robust features in our simulations that can be extrapolated to real stars or planets. We also revisit here the dependence of $\nu_E$ on tidal amplitude $a_0$. 
	
	For these simulations our initial conditions are small amplitude, solenoidal, homogeneous random noise for the velocity field and zero temperature fluctuation. We use a pseudo-random number generator seeded from the system clock in order to ensure that the initial conditions are unique in each simulation to a high probability.

	\section{Results}	\label{section_results}
	\subsection{Convection without shear}\label{section_convection_noshear}	
	\begin{figure*}
		\includegraphics[width=2\columnwidth]{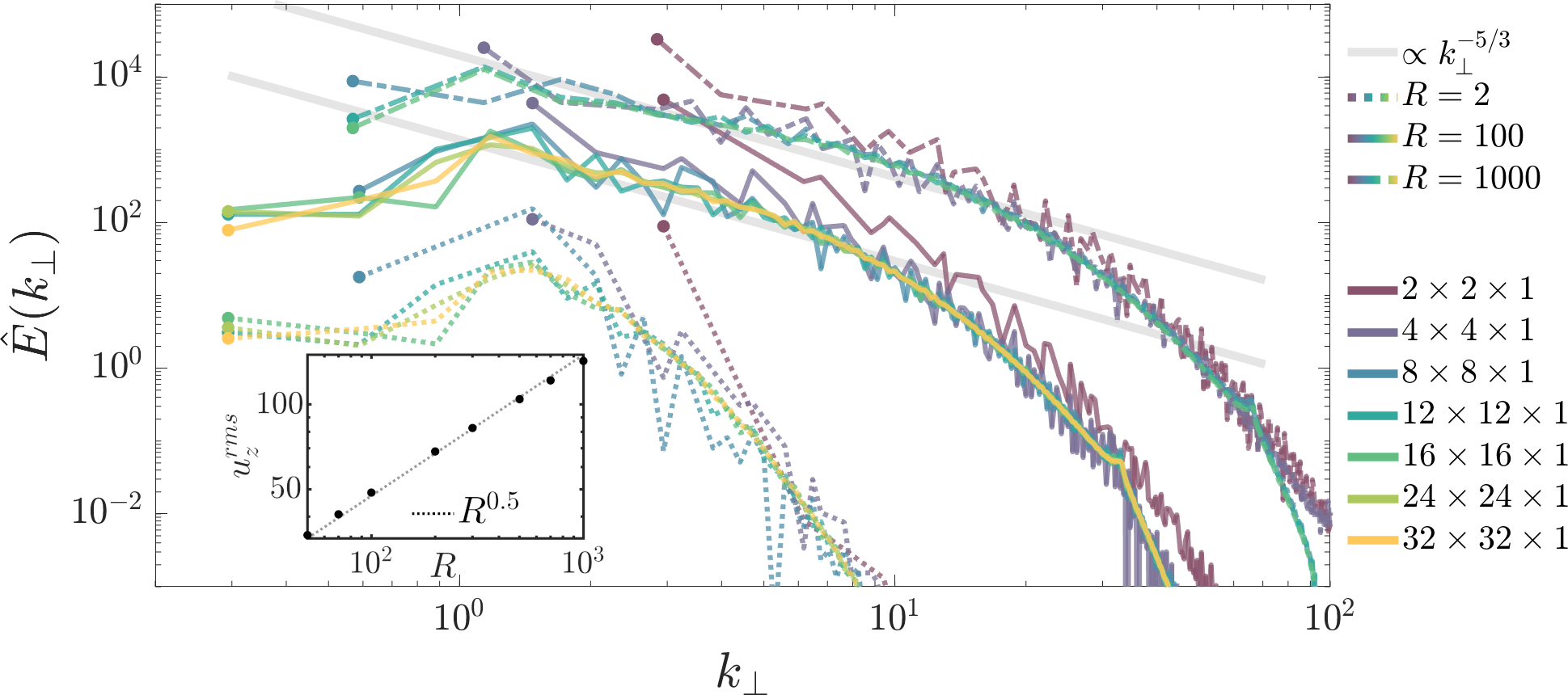}
		\caption{Time-averaged and vertically-integrated kinetic energy spectrum, $\hat{E}$, as a function of horizontal wavenumber $k_{\perp}$, for $R\in \{2,100, 1000\}$ (denoted by	dashed, solid and dot-dashed lines, respectively) and various domain sizes (see legend) for convection in the absence of oscillatory shear ($a_0 = 0$). 
		These spectra are visually indistinguishable (above the inherent variability within the convection) when a weak tidal shear is applied. The thick grey lines show the classical Kolmogorov $-5/3$ power law for the turbulent cascade of energy, for reference. The inset panel shows that the vertical convective velocities obey the classical diffusion-free mixing-length scaling \citep[e.g.][]{Spiegel1971}.      We note this data is for the fixed domain size of $(8,8,1)$ corresponding to the cases displayed in Fig.~\ref{figure_kinetic_energy_noshear}.     }
		\label{figure_boxes_resolution_spectra}
	\end{figure*}
	
	\begin{figure*}
		\includegraphics[width=2\columnwidth]{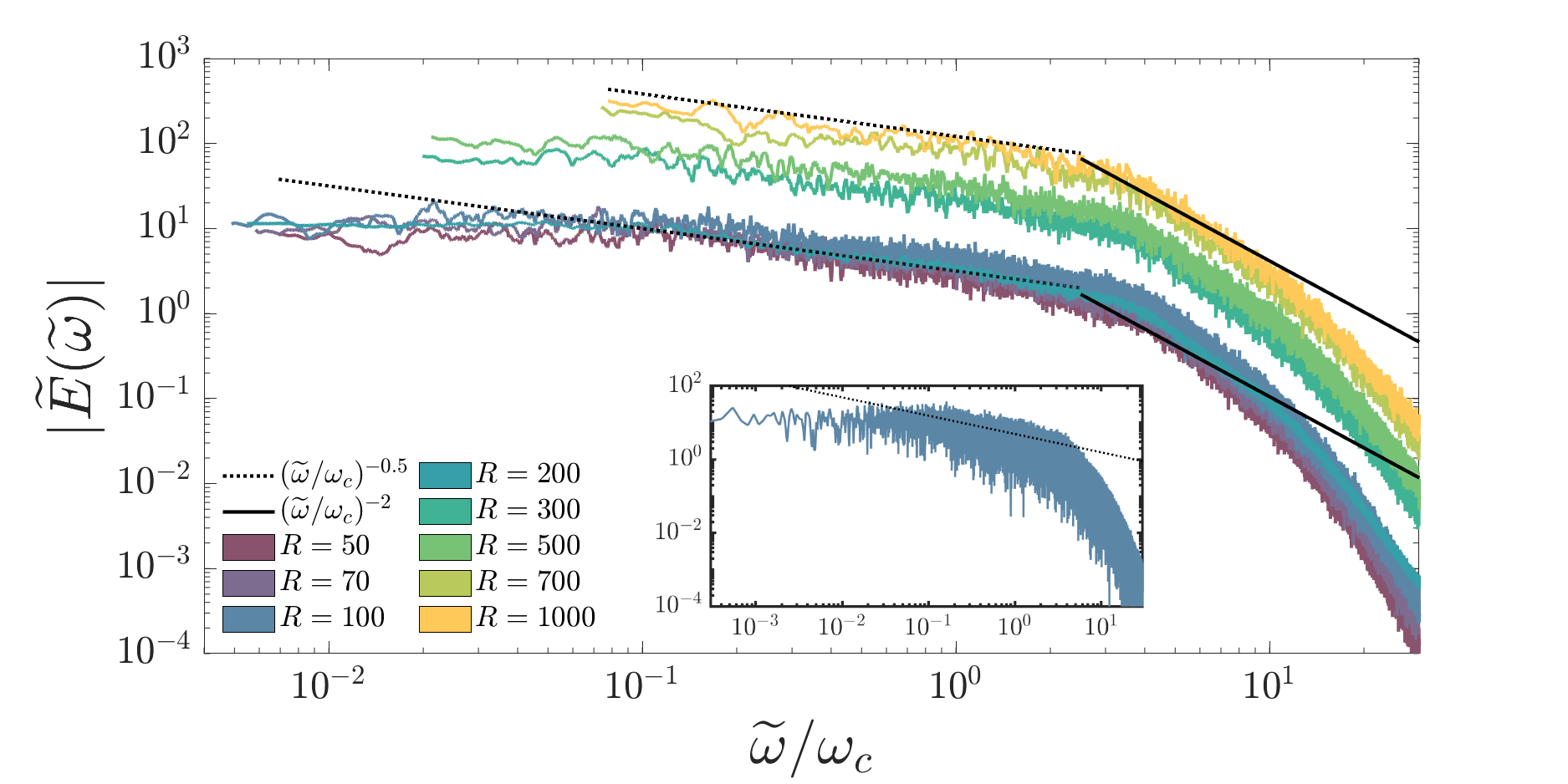}
		\caption{Frequency (temporal) spectra of the volume-averaged kinetic energy for various values of $R$ (see legend), with a domain size of $(8,8,1)$ in the absence of oscillatory shear. The solid lines highlight the scaling expected in the inertial range of $(\tilde{\omega}/\omega_c)^{-2}$ for a Kolmogorov cascade. For frequencies above those in the inertial range, we observe a power law decay of magnitude greater than $-2$ which can be attributed to the dissipation range (beyond which lies high-frequency noise). The dotted lines represent the $(\tilde{\omega}/\omega_c)^{-0.5}$ power law that exists at frequencies lower than the inertial range until a flattening of the spectrum at very low frequencies (corresponding with white noise). We note that the displayed spectra represent smoothed $20$-point moving averages of the full spectrum to reduce noise. The inset panel is an example of the frequency spectrum for the case of $R=100$ before smoothing.}
		\label{figure_kinetic_energy_noshear}
	\end{figure*}

	We begin our investigation by considering convection in the absence of oscillatory shear. In particular, for reasons that will become clear later on, we are interested in the wavenumber (eq.~\ref{numerical_1D_spectrum}) and frequency spectra of the energy (eq.~\ref{maths_freq_spectrum}) and how these vary as the strength of the convective driving $R$, and domain size $L_x=L_y$ are varied.

	We compute the wavenumber spectrum of the kinetic energy as described by eq.~\ref{numerical_1D_spectrum} for $R \in\{2,100,1000\}$ and various domain sizes $L_x = L_y \in\{2,4,8,12,16,24,32\}$. The results can be seen in Fig.~\ref{figure_boxes_resolution_spectra}. We ensure that the resolution per unit length for a given $R$ is held constant in all but the smallest domains, which are slightly better resolved. This ensures that the de-aliasing scale, defined by $k_{\perp}^{\text{alias}} = 2\pi N_x /(3L_x)$ (where $N_x$ is the number of grid-points in the $x$ direction), occurs for the same value of $k_{\perp}$. These simulations have been found to be well resolved by comparing the spectra for various resolutions. 
	
	From these results we observe an energetically dominant peak in the spectrum at $k_{\perp} \approx 2$ (which, we note, is similar to the onset wavenumber $\pi/\sqrt{2}$), which corresponds to a lengthscale of $L^{\text{peak}}\approx 3$, for each value of $R$. In cases with smaller $L_x$, the energy is instead preferentially dominated by the largest wavenumbers in the box. The spectrum in smaller domains is similar to those in larger domains except for the smallest $k_{\perp}$ values. 
	
	In the more turbulent cases with $R\in\{100,1000\}$, we can see from Fig.~\ref{figure_boxes_resolution_spectra} that they possess identifiable inertial ranges that extend from the peak of the spectrum until $k_{\perp} \approx (20,30)$, respectively, which are consistent with the classical Kolmogorov $-5/3$ power law \citep[e.g.][]{kolmogorov_local_1941,davidson_turbulence_2015}.	For even higher $k_{\perp}$, we observe a dissipation range in which the energy falls off faster with $k_\perp$. The cases with $R=2$ for all domain size are laminar and lack a clear inertial range. We show in the inset panel of Fig.~\ref{figure_boxes_resolution_spectra} that the convective velocities obey the classical diffusion-free scaling of mixing-length theory, such that $u_z^{\mathrm{rms}}\propto R^{0.5}$.
	
	We note that in order to make fair comparisons the spatial resolution has been chosen such that the aliasing scale takes the same value of $k_x\,,k_y$. The result of this is that in the larger domains there are more $k_x\,,k_y$ pairs to be distributed in the $k_{\perp}$ bins, which we note we have fixed for each $R$ case. As such the smaller domains have lower resolution in $k_{\perp}$ space than the larger domains despite being equivalently resolved in real space, hence the increased variation in the spectrum for smaller domains. Various statistics for these cases can be seen in Table~\ref{table_allsims_noshear} in Appendix \ref{appendixA}, which show that we attain convergence for sufficiently large $L_x=L_y$.

	We next compute the frequency spectrum of the kinetic energy, as described by eq.~\ref{maths_freq_spectrum}, for various cases with $R \in \{50, 70, 100, 200, 300, 500, 700, 1000  \}$ and a fixed domain size of $(8,8,1)$, which has been guided by our analysis of the wavenumber spectrum. These are shown in figure~\ref{figure_kinetic_energy_noshear}, which are computed by using a $20$-point moving average in order to smooth the original noisy signal (see insert). The angular frequency $\tilde{\omega}$ in each case has been scaled by the convective frequency.
	
	In the Kolmogorov description of turbulence, the inertial range follows a $-2$ power law in the frequency spectrum \citep{Landau1987Fluid,kumar_applicability_2018}. This power law is highlighted in figure~\ref{figure_kinetic_energy_noshear} by the solid black line. For each value of $R$, this inertial range begins at $\tilde{\omega}/\omega_c \approx 3$ and extends to higher frequencies with increasing $R$. In the case of $R=1000$ this inertial range extends until $\tilde{\omega}/\omega_c \approx 6$ while for $R=50$ the range is vanishingly small. This can more clearly be seen in Appendix \ref{appendixB} where Fig.~\ref{figure_kinetic_energy_noshear} has been re-plotted with application of an $(\tilde{\omega}/\omega_c)^2$ scaling factor (as well as zooming in on a narrower range of frequencies) which highlights the short inertial ranges. We observe a dissipation range above the inertial range, as is evident from the more rapid drop-off in the energy for higher frequencies. The key feature of this figure is our observation of a new power-law for intermediate frequencies $\tilde{\omega}/\omega_c \lesssim 3$, with an approximate exponent of $-0.5$ which extends over approximately two decades to lower frequencies. For very low frequencies, $\tilde{\omega}/\omega_c \lesssim 10^{-2}$, the spectrum then flattens off to indicate frequency-independent white noise. We note that not all of our spectra extend to low enough frequencies to observe the appearance of this white noise due to computational limitations.

\begin{figure}
		\includegraphics[width=\columnwidth]{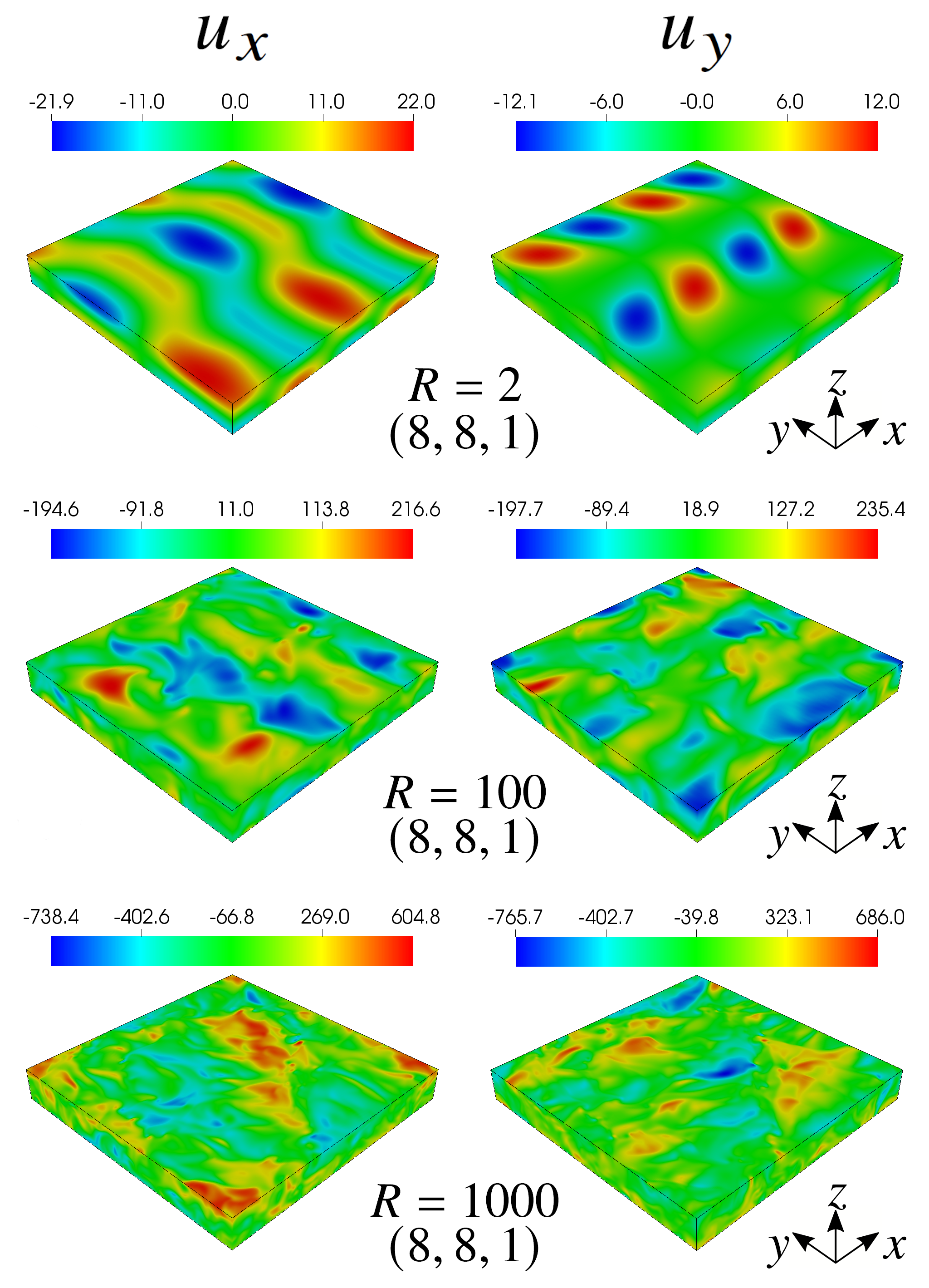}
		\caption{Snapshots of the $u_x$ (left) and $u_y$ (right) velocity components for convection in the absence of oscillatory shear. The values of $R\in\{ 2,100,1000\}$ and the domain size are displayed in each panel. As expected, as $R$ is increased, ever smaller scales are manifested in the flow (as seen in the bottom three rows), though large-scale components remain. The flow in the sheared cases is similar.}
		\label{figure_flow_plots}
	\end{figure}

	Snapshots of the horizontal flow showing the $u_x$ and $u_y$ components of velocity at chosen times are presented in Fig.~\ref{figure_flow_plots} for example simulations with $R \in\{2,100,1000\}$, all in $(8,8,1)$ domains. We note that these snapshots are also representative of cases including the oscillatory shear, since the flow is not strongly modified by its presence.

	In the more turbulent cases, $R\in\{100,1000\}$, the flow is fully three-dimensional and temporally chaotic for all domain sizes explored. As $R$ is increased ever smaller features in the flow appear, which is consistent with the extension of the inertial range in 
	Fig.~\ref{figure_boxes_resolution_spectra}. For the laminar cases with $R=2$ the flow consists of spatially persistent features with temporally-periodic amplitudes that are similar to the results in smaller domains presented in \citetalias{duguid_tidal_2019}. The frequency spectrum for this laminar case, and in smaller domains, consists of discrete peaks. On the other hand, we comment that $R=2$ simulations in larger domains with $L_x = L_y \geq 12$ instead exhibit a chaotic flow (which is still non-turbulent due to the lack of an inertial range), with a frequency spectrum that is more similar to those with larger $R$ values.

	\subsection{Frequency dependence of the effective viscosity}
	\begin{figure*}
		\includegraphics[width=1.8\columnwidth]{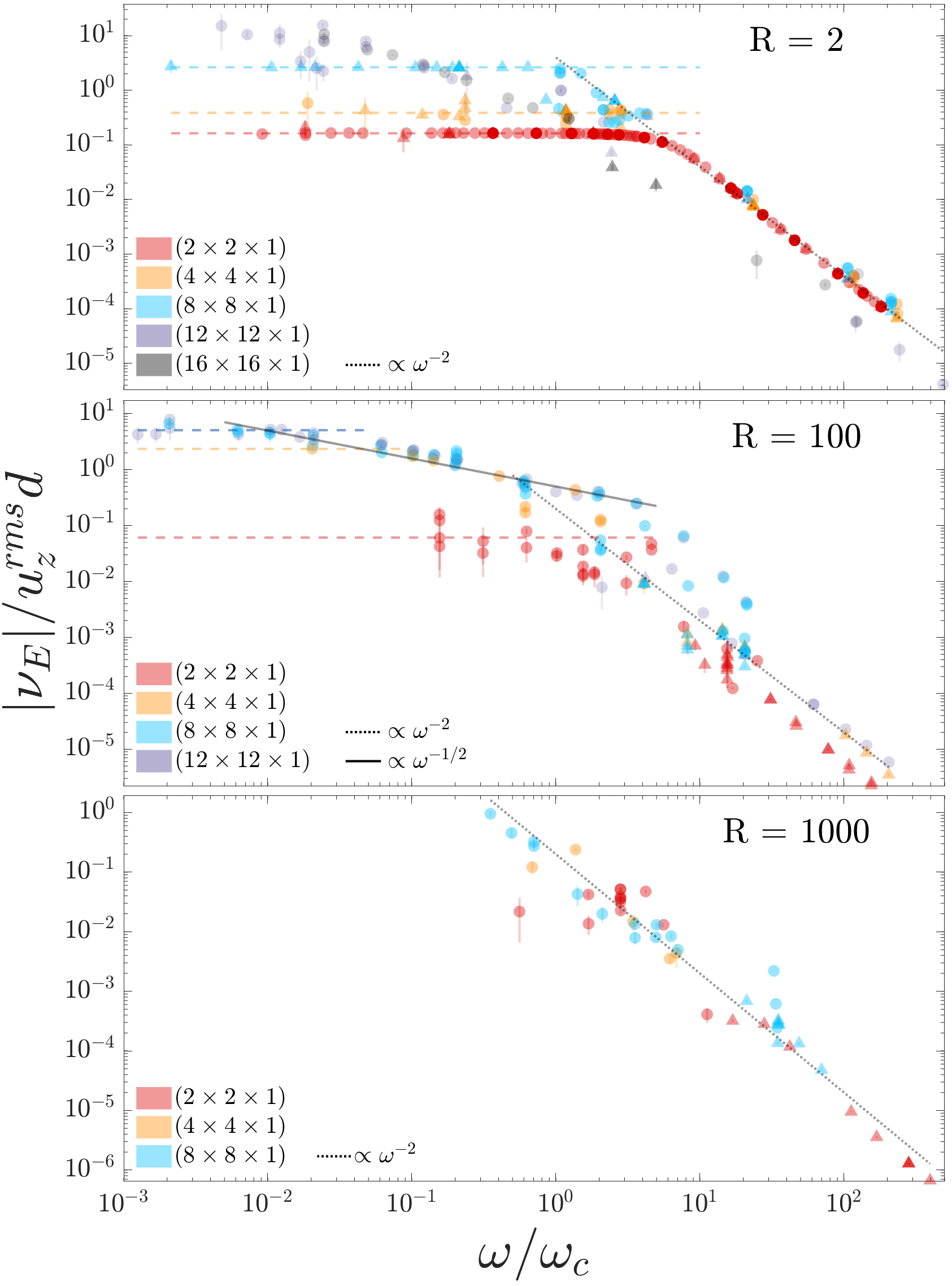}
		\caption{Scaled effective viscosity $\lvert\nu_E \rvert/u_z^{\text{rms}} d$ as a function of scaled shear frequency, $\omega/\omega_c$, that arises from the interaction between the oscillatory tidal flow and convection. Various shear amplitudes are shown, in the range $a_0 \in[0.05,1]$, and the effects of $a_0$ will be discussed later (and shown in Fig.~\ref{figure_nu_with_spectra}). The cases shown have $R\in\{2,100,1000\}$ (top to bottom respectively) with various domain sizes (see legend). We denote the sign of the effective viscosity using circles for positive and triangles for negative values. Error bars are shown but these are often smaller than the sizes of the symbols. The dashed lines show horizontal fits to the low frequency regime. The solid lines show the $(\omega/\omega_c)^{-0.5}$ scaling for intermediate frequencies, which is a new result in this work. The dotted lines show the classic $(\omega/\omega_c)^{-2}$ scaling for high-frequency tidal forcing. The scatter of points in the frequency range $\omega/\omega_c = (10^{0}, 10^{2})$ for $R=100$ and $\omega/\omega_c \approx 40$ for $R=1000$ can be attributed to a shear amplitude dependence that is shown more clearly in a later in Fig.~\ref{figure_nu_with_spectra}. }
		\label{figure_alpha_plots}
	\end{figure*}
	
	We now move on to the main task of the paper, which is to analyse the interaction between oscillatory tidal (shear) flows and convection.
	The oscillatory (tidal) shear flow, described by eq.~\ref{maths_shear}, is now introduced and we begin by presenting results for the magnitude of the scaled effective viscosity $\alpha$ (eq.~\ref{maths_mlt}) in simulations with various values of $R\in\{2,100,1000\}$, $\omega = [0.001,10000]$, $L_x(=L_y) \in\{2,4,8,12,16\}$ and $a_0 = [0.05,1]$. Details of the typical ranges of these parameters for various cases, including further details such as the resolution, are given in Table~\ref{table_allsims} and appendix~\ref{appendixA}. The results are presented in Fig.~\ref{figure_alpha_plots}, where $\omega$ has been scaled by the relevant convective frequency $\omega_c$. The main result here is that $\nu_E$ is a frequency-dependent quantity and is strongly attenuated for high-frequency tidal forcing, in agreement with prior works (\citealt{ogilvie_interaction_2012}; \citetalias{duguid_tidal_2019}; \citealt{vidal_turbulent_2020}).
	
	In order to reduce the influence of noise on the computation of $\nu_E$, we evaluate eq.~\ref{maths_eff_visc} by computing the cumulative integral, to which we apply a linear fit to determine $\nu_E$. This method also allows us to define an error in $\nu_E$ using two standard deviations from the mean slope, as adopted in \citetalias{duguid_tidal_2019}.	To ensure convergence, the simulations are run for tens to thousands of tidal periods (with the exception of some cases with $\omega<0.1$ that could only be run for approximately one tidal period). These long-duration simulations were successful in reducing the error in the computation of $\nu_E$, which is demonstrated by the small error bars in Fig.~\ref{figure_alpha_plots}, which are typically smaller than the symbols plotted. 
	
	In the high frequency regime $\omega/\omega_c \gtrsim 5$, for all values of $R$, we observe a clear $-2$ power law (represented by the dotted lines in Fig.~\ref{figure_alpha_plots}). This corresponds with the high-frequency scaling law ($\nu_E\propto \omega^{-2}$) of \cite{goldreich_turbulent_1977}, and clearly disagrees with the high-frequency scaling law ($\nu_E\propto \omega^{-1}$) of \citep{zahn_les_1966}. This result is consistent \citetalias{duguid_tidal_2019}, as well as prior simulations of similar problems such as homogeneous convection \citep{ogilvie_interaction_2012} and ABC flows \citep{braviner_stellar_2015}. The theory of \cite{goldreich_turbulent_1977} assumes a Kolmogorov turbulent cascade to obtain a $-2$ power law for $\nu_E$.
	In \citetalias{duguid_tidal_2019} we noted that $R=2$ simulations were laminar and yet still followed the $-2$ scaling. This remains true for the larger domains considered here, thus demonstrating that a turbulent flow is not required to obtain a $-2$ power-law scaling for $\nu_E$ at high frequencies. The behaviour of laminar convection with $R=2$ can probably be explained by applying the asymptotic theory developed in \citetalias{duguid_tidal_2019}, which extends that of \cite{ogilvie_interaction_2012}, providing all convective modes are accounted for, though we do not attempt to do so here as our larger domains would require considering many modes. We will later show (see Fig.~\ref{figure_nu_with_spectra}) that the scatter in the high frequency regime for $R=100$ can be attributed to an amplitude ($a_0$) dependence of $\alpha$.

	Fig.~\ref{figure_alpha_plots} provides evidence for a previously undiscovered scaling $\nu_E\propto \omega^{-0.5}$ for intermediate frequencies with $\omega/\omega_c \approx (10^{-2},1)$. This new regime is clearly observed in the middle panel with $R=100$ in all domains with $L_x>2$, and is highlighted by the solid line representing a {-0.5} power law (this regime is also present with $R=2$ in the largest domains $L_x \geq12$). To the best of our knowledge, this is the first time this scaling has been observed in simulations, and it has also never been predicted theoretically. The cases with $L_x=2$ previously presented in \citetalias{duguid_tidal_2019} instead exhibit a frequency-independent $\nu_E$ for $\omega\lesssim \omega_c$. This difference demonstrates the importance of resolving the peak of the spatial spectrum (see Fig.~\ref{figure_boxes_resolution_spectra}). This new intermediate regime is not evident in the $R=1000$ simulations, probably because we have not been able to run simulations for sufficiently low frequencies to observe it clearly (these cases are particularly computationally expensive). The lowest frequencies for $R=1000$ may be starting to transition to this regime, but we are unable to confirm this at present.

	For cases exhibiting an intermediate regime with a $-0.5$ power law, the magnitude of $\alpha$ becomes independent of the domain size and tidal amplitude, as long as the domain size is large enough to resolve the peak	of the wavenumber spectrum (Fig.~\ref{figure_boxes_resolution_spectra}). We also note that the magnitudes of $\nu_E$ in the intermediate and low frequency regimes are significantly larger than for the cases in smaller boxes. They are also larger, by more than an order of magnitude, from the naive expectation from MLT, which would predict $\alpha=1/3$ \citep{Zahn1989}. This suggests that convection is more efficient at damping low frequency tidal flows than previously expected.

	The $R=2$ case exhibits a change in behaviour in this intermediate frequency range as we increase the domain size, from frequency-independent behaviour in smaller boxes, to following a $-0.5$ power law in larger boxes. This coincides with our observation that the flow transitions from deterministic to chaotic in the largest boxes, as well as being related to the requirement to resolve the energetically dominant scales (see Fig.~\ref{figure_boxes_resolution_spectra}), which we will address further in section~\ref{section_frequency_visc}. Note that the flow is non-turbulent for $R=2$, and yet it still exhibits the same $-0.5$ scaling for $\nu_E$.

	In \citetalias{duguid_tidal_2019} we observed a frequency independent regime for $\omega/\omega_c \lesssim 5$, which can be seen in Fig~\ref{figure_alpha_plots}, in domain sizes of $(2,2,1)$ for all $R$ (it also occurs in domains up to $L_x=8$ for the $R=2$ cases). In larger domains, this frequency-independent regime is only observed for very low frequencies, $\omega/\omega_c \lesssim 10^{-2}$. We have only observed this regime for $R=100$ due to the computational expense of probing such low values of $\omega$. Indeed, these typically require approximately $1000$ diffusion times to obtain convergence in the evaluation of $\nu_E$. Where possible, we have shown the best fit to the frequency-independent regime with dashed lines in Fig.~\ref{figure_alpha_plots}.

	On physical grounds, there are no restrictions on the sign of the effective viscosity defined by eq.~\ref{maths_eff_visc}. Indeed, in \citetalias{duguid_tidal_2019}, as suggested in the earlier simulations of \cite{ogilvie_interaction_2012}, we observed robust negative values for $\nu_E$ for very high frequencies in the turbulent cases. In Fig.~\ref{figure_alpha_plots} we have denoted positive values with circles and negative by triangles. In \citetalias{duguid_tidal_2019}, we found that in laminar cases the initial conditions determined the sign of $\nu_E$.	This behaviour is again observed in the $R=2$ cases up to domain sizes of $(8,8,1)$. In the $(12,12,1)$ cases, where the flow is chaotic, and there is an increase of energy transfer between convective modes, there appears to be a preference towards positive values for $\nu_E$, with negative values only occurring around the transition between the intermediate and high frequency regimes.
	
	For the more turbulent $R\in\{100,1000\}$ cases with the domain size $(8,8,1)$ we observe the same behaviour as in the smaller box simulations of \citetalias{duguid_tidal_2019}, in that $\nu_E$ transitions from positive ($\omega/\omega_c \lesssim 10$) to negative values ($\omega/\omega_c \gtrsim 10$). However we note that for the $R=100$ cases in the large domain $(12,12,1)$, $\nu_E$ is also positive for frequency ratios much larger than 10. We have also conducted simulations in small $(2,2,1)$ domains with $R=10000$ (not presented) where the transition to negative values is shifted to higher frequencies, $\omega/\omega_c \approx 30$, than for $R\in\{100,1000\}$. This suggests that the transition to negative values occurs for unrealistically high tidal frequencies in convection with astrophysically-relevant values of $R$.

	\begin{figure*}
		\includegraphics[width=2\columnwidth]{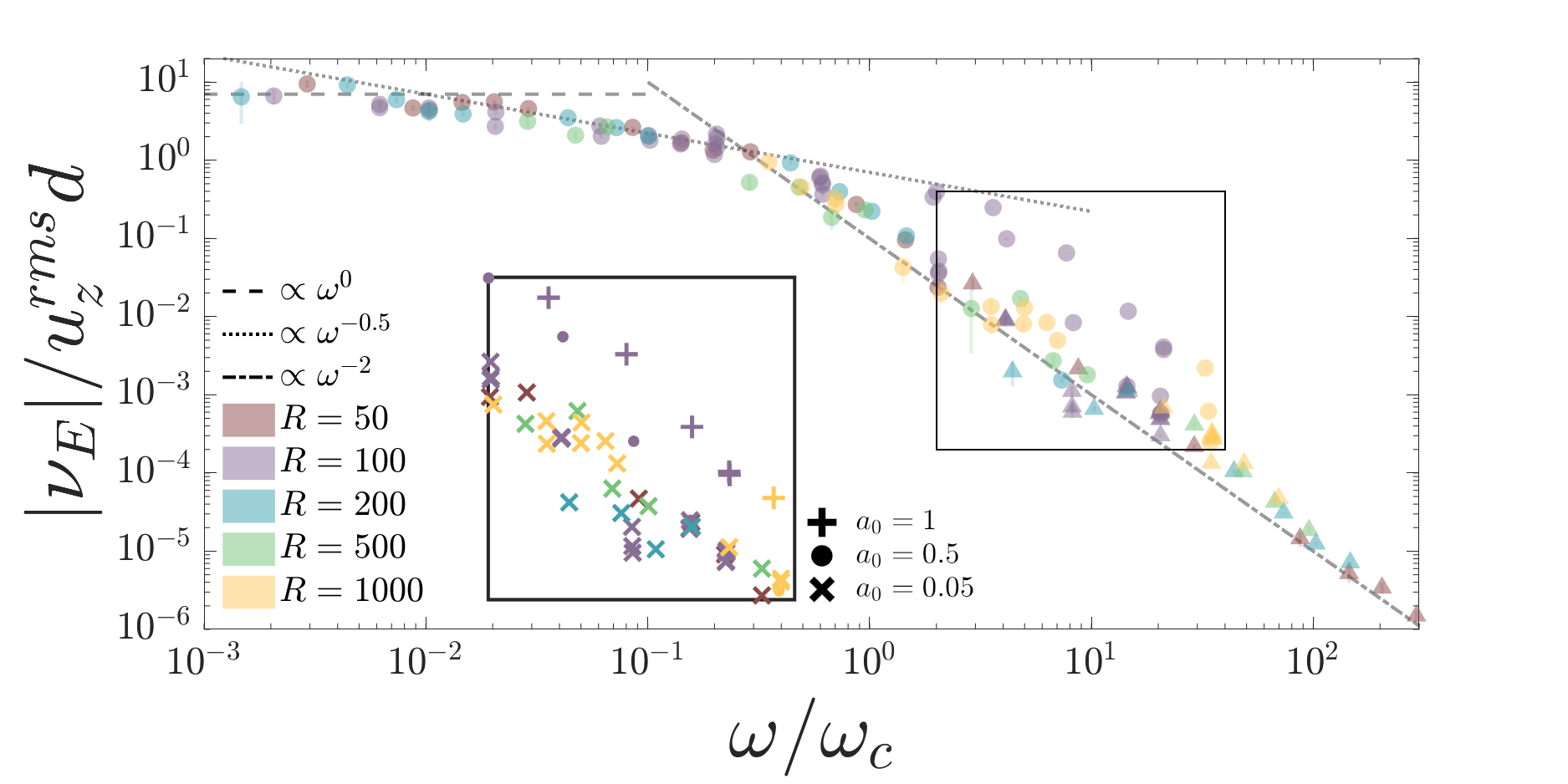}
		\caption{Scaled effective viscosity as a function of the scaled shear frequency, similar to figure~\ref{figure_alpha_plots}. The cases shown here have $R\in\{50,100,200,500,1000\}$ (see legend) for a fixed domain size of $(8,8,1)$ and with shear amplitudes in the range $a_0 = [0.05,1]$ (not highlighted here to avoid confusion).  Power law lines are displayed for exponents of $\{0,-0.5,-2\}$ for low, intermediate and high frequency cases respectively (see legend). This shows that the frequency-dependence of the effective viscosity is robust to changes in the Rayleigh number over this range. The scatter of points in the frequency range $\omega/\omega_c = (10^{0}, 10^{2})$ can be attributed to a shear amplitude dependence that is illustrated clearly in the inset panel.}
		\label{figure_alpha_many_R}
	\end{figure*}
	
	Convection in stars is much more turbulent, with much larger values of $R$, than we can explore in our simulations. One of our key goals is to determine if there are robust features or scaling laws as $R$ is varied. In Fig.~\ref{figure_alpha_many_R}, we compare the scaled effective viscosity as a function of the scaled shear frequency for various values of $R\in\{50,100,200,500,1000\}$ in a fixed domain size $(8,8,1)$. 
	
	Fig.~\ref{figure_alpha_many_R} shows the frequency dependence of $\nu_E$ in our simulations. We include a range of $R\geq50$ in this plot, but by plotting $|\nu_E|/u_z^{rms} d$ we collapse the low frequency data ($\omega/\omega_c < 1$) onto a single \lq \lq master curve". This collapse is only observed in simulations in large domains that resolve the energetically-dominant lengthscale, for which further increases in domain size are not observed to strongly affect our results (see Fig.~\ref{figure_alpha_plots}). The scaling adopted here assumes the convection to approach a diffusion-free mixing-length regime in which convective velocities scale as $\sqrt{R}$ (which is demonstrated in the inset panel in Fig.~\ref{figure_boxes_resolution_spectra}; e.g.~\citealt{Spiegel1971}), such that $\omega_c$ and $\nu_E$ would also be expected to scale as $\sqrt{R}$. We have therefore demonstrated in Fig.~\ref{figure_alpha_many_R} that the convective velocities (and dominant length-scales) for $R\geq50$ are essentially in the diffusion-free mixing-length regime for our simulations. At higher frequencies, there is still considerable scatter which
	comes mainly from dependence on the shear amplitude $a_0$. In the inset, symbols $+$ correspond to $a_0=1$, $\bullet$ to $a_0=0.5$ and $\times$ to $a_0=0.05$. There is a systematic increase in $\nu_E$ with amplitude,
	which we will explore further below. However, the data for the same values of $a_0$ collapse reasonably well onto their $R$-independent curves even for high frequency tides. For numerical reasons it is difficult to get consistent results for $\nu_E$ with low amplitude tides at low frequencies, so the low frequency points $\omega/\omega_c < 0.1$ are mostly for amplitudes above 0.5, but generally we found little evidence for significant variation of scaled effective viscosity with amplitude at low frequency.
	
	The frequency-independent low frequency regime is observed when $\omega/\omega_c < 10^{-2}$, for each of $R=50$, $100$ and $200$, though we should point out that there only are few simulations with such low frequencies. The dashed line in this case is the linear fit to the constant slope for the $R=100$ cases, which also matches those with $R=50$ and $200$. 
	
	It is clear from Fig.~\ref{figure_alpha_many_R} that the new intermediate scaling regime for $\omega/\omega_c \approx (10^{-2}, 1-5)$ holds for all $R$ values explored, highlighting that this new regime is also robust. In the high frequency regime, when $\omega/\omega_c \gtrsim 1-5$, we observe a robust transition to a $-2$ power law for all values of $R$. However, in this regime, there is more scatter in the points from a single ``master curve", which can be attributed a shear-amplitude dependence of our results, since here we adopt various $a_0 \in\{0.05, 0.5, 1\}$ (as we will explain further below).
	
	Negative (positive) values of the effective viscosity in Fig.~\ref{figure_alpha_many_R} are denoted by triangles (circles). If we consider the lowest frequency for each $R\in\{50,100,200,500,1000\}$ for which $\nu_E$ is negative, we find this to occur at approximately $\omega/\omega_c \in \{2, 4,5, 30,  20 \}$, respectively. This shows that with exception of the $R= 500$ case, the larger the value of $R$ the higher frequency required to obtain negative values of the effective viscosity. We also note that the apparent discrepancy in the $R=500$ case could be the result of the particular discrete values of the frequency that have been run.

	\subsection{Comparing the frequency spectra to effective viscosity}\label{section_frequency_visc}
	
\begin{figure}
		\includegraphics[width=\columnwidth]{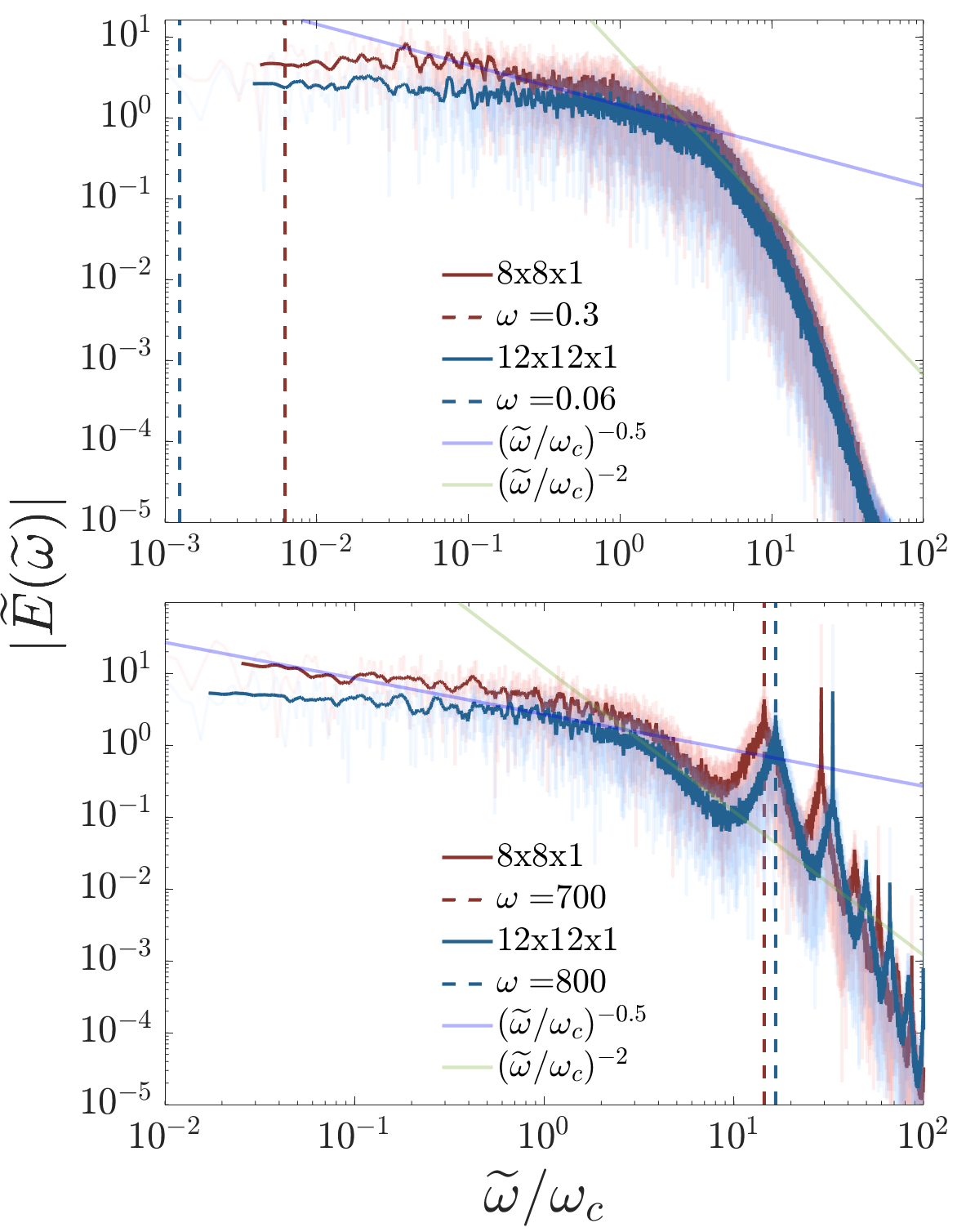}
		\caption{Examples of the frequency spectrum of the kinetic energy  for convection in the presence of oscillatory shear, with $R=100$, $a_0 = 0.5$, $(L_x,L_y,L_z) \in \{(8,8,1), (12,12,1)\}$ (red and blue, respectively) and $\omega \in \{0.3, 0.06\}$ (top, with $\omega<\omega_c$) and $\{700,800\}$ (bottom, with $\omega > \omega_c$). The angular frequency has been scaled by the relevant convective frequency. The solid lines show the 20-point moving average of the full spectrum, which is shown using faded lines. An inertial range is observed, where the spectrum follows an $(\tilde{\omega}/\omega_c)^{-2}$ power law (green line), before entering the dissipation range for the highest $\tilde{\omega}$. We also observe a significant region which features a $(\tilde{\omega}/\omega_c)^{-0.5}$ power law in the spectrum, which matches the scaling observed for $\nu_E$ in Figs.~\ref{figure_alpha_plots} and \ref{figure_alpha_many_R}. The vertical dashed lines correspond to the frequency of the oscillatory shear in each case. Similar results are obtained for the frequency spectrum of the Reynolds stress. }
		\label{figure_kinetic_energy_spectum}
	\end{figure}

	In previous work (\citealt{penev_direct_2009}; \citetalias{duguid_tidal_2019}; \citealt{vidal_turbulent_2020}) it has been suggested that the frequency (temporal) spectrum of the kinetic energy $\tilde{E}(\tilde{\omega})$ (or Reynolds stress) may play an important role in determining the frequency-dependence of the effective viscosity. In this section we explore more closely the connection between the frequency spectrum of the kinetic energy and the frequency dependence of the effective viscosity. Examples of the frequency spectrum of the kinetic energy (evaluated after applying a Hann window function) can be seen in Fig.~\ref{figure_kinetic_energy_spectum} for the low frequency tide regime with $\omega \in \{ 0.3,0.06\}$ (top) and the high frequency tide regime with $\omega \in \{ 700,800\}$ (bottom). All cases in Fig.~\ref{figure_kinetic_energy_spectum} have $R=100$ and a shear amplitude $a_0 = 0.5$ covering two cases in each of the high and low frequency regimes with domain sizes $L_x \in\{8,12\}$ (dark red and dark blue, respectively) and the tidal frequencies (see legend) denoted by the vertical dashed lines. For these plots we scale the angular frequency $\tilde{\omega} = 2\pi/\tau$ (where $\tau$ represents the period of each Fourier component) of the spectrum by the relevant convective frequency ($\omega_c$) in each simulation, so as to make a meaningful comparison between these spectra results and the profiles of $\nu_E$. In Fig.~\ref{figure_kinetic_energy_spectum} the solid lines represent the $20$-point moving average of each full kinetic energy spectrum (which are plotted using faded lines). We also note that similar spectra have been obtained for all simulations that have been run for a long enough duration.
	
	In these low tidal frequency example cases (Fig.~\ref{figure_kinetic_energy_spectum} top) we observe a small inertial-like range defined by a $-2$ power law (green line) \citep{Landau1987Fluid,kumar_applicability_2018} in the kinetic energy spectra starting at $\tilde{\omega}/\omega_c \approx 4$ and extending to $\tilde{\omega}/\omega_c \approx 10$. Beyond this, we observe a dissipation range where the spectrum transitions to a power law decay with magnitude greater than $2$, followed by low-power noise at very high frequencies as a consequence of the finite time-step size. For frequencies lower than the inertial range, we observe a power law exponent that is consistent with $-0.5$ (light blue line) extending down to $\tilde{\omega}/\omega_c \approx 10^{-1}$ before the spectrum approaches white noise for the lowest observable frequencies. The frequency spectrum of the kinetic energy and Reynolds stress are both consistent with the un-sheared cases presented in section~\ref{section_convection_noshear} for all cases where the shear frequency is in the low or intermediate regimes. We note that in the high frequency regime the shear introduces a strong resonant response in the spectrum at the shear frequency, which is related to the larger shear amplitude $a_0\omega$ at high frequencies. The rapid drop-off in the frequency spectrum then allows the energy injected by the shear to become observable for these high frequencies. Fig.~\ref{figure_kinetic_energy_spectum} also shows that the shape of the frequency spectrum is independent of domain size, providing the energetically-dominant convective modes are resolved spatially.
	
	In the high frequency cases (Fig.~\ref{figure_kinetic_energy_spectum} bottom) the spectrum behaves similarly to the low frequency cases when $\tilde{\omega}/\omega_c \lesssim 5$. We observe a significant modification of the spectrum in the high frequency regime beginning with a substantial peak in the spectrum at the shear frequency. The peak is not confined to the discrete frequency of the shear and has a substantial lead and lagging tail. Further, we observe a significant resonant chain of peaks each with the same shape as the main peak. 
	
\begin{figure*}
		\includegraphics[width=1.8\columnwidth]{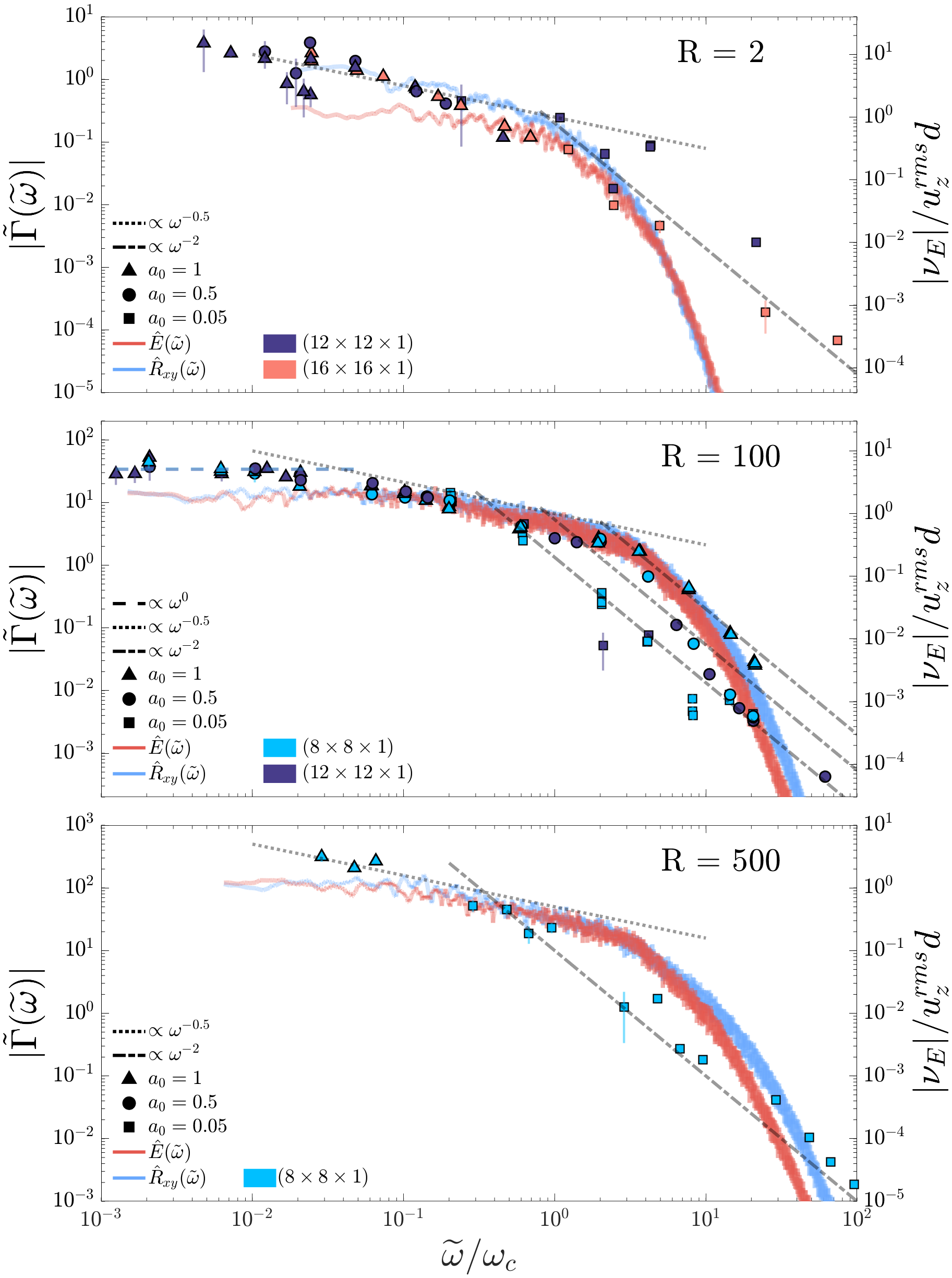}
		\caption{Magnitude of the effective viscosity from figure~\ref{figure_alpha_plots} (right axis - points) with the frequency spectra of the convection over-plotted, $\tilde{\Gamma}(\tilde{\omega})$ (left axis - lines) where $\Gamma \in\{E,R_{xy}\}$ is either the kinetic energy, $E$, (blue) or Reynolds stress, $R_{xy}$, (red). The $R=1000$ case has been replaced with $R=500$ due to computational limitations of reaching the lowest frequencies with higher $R$ values. The symbols for the effective viscosity now highlight the amplitude $a_0$ for each case, while the sign is no longer identified in this figure to clarify the presentation. The symbol colour denotes the domain size (see legend). The 20-point moving average of the frequency spectrum is shown here. Various relevant power laws are denoted by the dotted and dashed lines which are highlighted in the legend. } 
		\label{figure_nu_with_spectra}
	\end{figure*}

	We now compare the frequency spectra (evaluated after applying a Hann window) of kinetic energy (blue) and Reynolds stress (red) with the scaled effective viscosity in figure~\ref{figure_nu_with_spectra}. 
	Note that the symbols denoting the scaled effective viscosity
	are shown as a function of the scaled tidal frequency, $\omega/\omega_c$, whereas the spectrum is plotted as a function of $\tilde{\omega}/\omega_c$. As before, the solid lines for the spectrum represent its 20-point moving average. We demonstrate these comparisons for cases similar to those in figure~\ref{figure_alpha_plots}, that is, cases with $R\in\{2,100,500\}$ (cases with $R=1000$ were excluded due to the difficulty in reaching the intermediate regime), domain sizes $L_x=L_y\in\{2,4,8,12,16\}$ and various amplitudes $a_0 = [0.05,1]$. For each value of $R$, we show a representative spectrum of kinetic energy and Reynolds stress, choosing cases with the longest run time to effectively probe the low frequency regime. As the domain size does not significantly alter the spectrum as long as it is ``large enough" (see figure~\ref{figure_kinetic_energy_spectum}), we plot a case with a domain size $(8,8,1)$ for each $R$. Similarly, the shear in the low-frequency regime only weakly affects the spectrum, as we can observe from comparing figures~\ref{figure_kinetic_energy_noshear} and \ref{figure_kinetic_energy_spectum}, so we adopt a representative case for each $R$ with $\omega < 1$ that has the longest run time. In figure~\ref{figure_nu_with_spectra} the symbols now denote the amplitude of the shear (see legend for the values).

	The key result of figure~\ref{figure_nu_with_spectra} is that for low and intermediate frequencies such that $\omega/\omega_c\lesssim 5$, the frequency dependence of $\nu_E(\omega)$ closely follows the spectrum of the energy and Reynolds stress. This agrees with the global simulations of \cite{vidal_turbulent_2020}. It is an important result because it suggests that we can infer the frequency dependence of $\nu_E$ in stars if we know the spectrum of the convection.
	
	It is worth highlighting that, although the left and right $y$-axis values are offset, the range of values in both is similar. We also note that there is a good agreement between the spectrum of the kinetic energy and the Reynolds stress, though the kinetic energy has slightly smaller amplitude than the Reynolds stress.
	
	We continue our analysis of figure~\ref{figure_nu_with_spectra} by considering the high frequency regime where $\omega/\omega_c \gtrsim 1$. As alluded to earlier, we observe an amplitude dependence in the magnitude of $\nu_E$, which is most clearly observed in cases with $R=100$ and shear amplitudes of $a_0 \in\{0.05,0.5,1\}$. This amplitude dependence shifts where the transition to the $-2$ power law begins, which is here observed to occur when $\omega/\omega_c \approx (0.6,1.5,3)$, respectively. In the high frequency regime, $\nu_E\propto \omega^{-2}$ for higher $\omega$. This only agrees with the spectrum for a narrow range of frequencies corresponding to the inertial-like range. For higher frequencies, the spectrum transitions into a dissipation range, where the power law exponent is steeper than  $-2$, whereas the effective viscosity continues to follow the $-2$ power law. This again suggests that, despite the power law of the inertial range and the frequency dependence of the effective viscosity being the same, this cannot explain the robustness of $\nu_E\propto \omega^{-2}$ for high frequency tidal forcing.
	
	For intermediate frequencies, $\omega/\omega_c \approx (10^{-2},10^0)$, we observe a strong agreement in the power law of the effective viscosity with both the kinetic energy and Reynolds stress frequency spectra for all $R$ plotted. The transition from intermediate to high frequency regimes in the effective viscosity does not always coincide with when the spectrum falls off more steeply than a $-0.5$ power law. In fact, the $R=100$ cases clearly demonstrate that the amplitude dependence plays a role in deciding when the effective viscosity transitions to the quadratic scaling regardless of the slope of the spectrum. 
	
	Although we have shown that there is good agreement with the frequency-dependence of the scaled effective viscosity and the frequency spectrum of kinetic energy (or Reynolds stress) in the intermediate and low frequency regimes, the relationship between these quantities is a constant of proportionality. That is, we have shown that $\nu_E(\omega/\omega_c) \propto \tilde{E}(\tilde{\omega}/\omega_c)$. Since the intermediate and low frequency spectrum appears to be approximately amplitude and domain size independent, this constant of proportionality may be some function of $R$ (and possibly ${Pr}$ which we have not explored in this work). 
	
	For the $R=2$ cases shown in figure~\ref{figure_nu_with_spectra}, we only display domain sizes of $L_x=L_y\in\{12,16\}$, since these are required to obtain a $-0.5$ power law in the intermediate regime. This may be related to the transition to a chaotic flow and/or the requirement of resolving the energetically dominant scales in larger domains. Cases in smaller domains exhibit a frequency spectrum consisting of discrete peaks, suggesting little energy exchange between eddies with different time-scales. However, in the larger domains the frequency spectrum is more continuous. This suggests that this new $-0.5$ power law regime is a consequence of the frequency spectrum of chaotic/turbulent flow. In addition, the robustness of this new regime for both laminar and turbulent flows indicates that it may be relevant for understanding the interaction between tidal flows and convection in stars and giant planets.

	\begin{table*}
		\begin{tabular}{ c | c | c | c | c | c | c | c  }
			$R$     & $L_x (=L_y)$ & $\omega$  &  $\alpha = \overline{\left(\frac{\nu_E}{u_z^{\text{rms}}d} \right)}$ & $\gamma = \overline{\nu_E (\omega/\omega_c)^2}$   & $u_z^{\text{rms}}=d\omega_c$ & \begin{tabular}[c]{@{}c@{}}typical resolution\\$(N_x,N_y,N_z)$\end{tabular}  & $a_0$  \\ \hline\hline
			$2     $&$ 2     $&$ [0.001, 1000]   $&$ 0.163 $&$ 21.071 $&$ 5.484 $& $(64,64,32)$        &$ [0.0005 , 1] $                      \\
			$2     $&$ 4     $&$ [0.08, 1000]    $&$ 0.386 $&$ 20.538 $&$ 4.354 $& $(64,64,32)$        &$ 0.05       $                         \\
			$2     $&$ 8     $&$ [0.01, 1000]    $&$ 2.63  $&$ 25.629 $&$ 4.698 $& $(64,64,32)$        &$ 0.05      $                          \\
			$2     $&$ 12    $&$ [0.01, 2000]    $&$ -     $&$ 3.923  $&$ 4.278 $& $(128,128,64)$      &$ [0.05 , 1]  $                        \\
			$2     $&$ 16    $&$ [0.1, 500]      $&$ -     $&$ 4.888  $&$ 4.091 $& $(128,128,64)$      &$ [0.05 , 1]  $                        \\ \hline 
			$5     $&$ 2     $&$ [0.001, 1000]   $&$ 0.168 $&$ 23.663 $&$ 13.44 $& $(64,64,32)$        &$ 0.05      $                          \\\hline
			$10    $&$ 2     $&$ [2, 5000]       $&$ 0.144 $&$ 28.450 $&$ 22.87 $& $(64,64,128)$       &$ 0.05       $                         \\ \hline
			$50    $&$ 8     $&$ [0.1, 10000]    $&$ -     $&$ 5.761  $&$ 34.39 $& $(128,128,128)$     &$ [0.05 , 1] $                         \\\hline
			$100   $&$ 2     $&$ [10, 10000]     $&$ 0.061 $&$ 3.967  $&$ 63.85 $& $(64,64,128)$       &$ [0.0005 , 1]$                        \\
			$100   $&$ 4     $&$ [1, 10000]      $&$ 2.356 $&$ 8.318  $&$ 49.07 $& $(64,64,128)$       &$ [0.05 , 1]   $                       \\
			$100   $&$ 8     $&$ [0.05, 1000]    $&$ 5.109 $&  see table~\ref{table_amp_depend}     &$ 49.06 $& $(128,128,128)$                  &$ [0.05 , 1]       $                   \\
			$100   $&$ 12    $&$ [0.06, 10000]   $&$ 5.074 $&$ 11.794 $&$ 48.78 $& $(128,128,128)$     &$ [0.05 , 1]    $                      \\\hline
			$200   $&$ 8     $&$ [0.1, 10000]    $&$ -     $&$ 11.177 $&$ 68.16 $& $(128,128,128)$     &$ [0.05 , 1]     $                     \\\hline
			$500   $&$ 8     $&$ [30, 10000]     $&$ -     $&$ 24.785 $&$ 104.1 $& $(256,256,192)$     &$ 0.05            $                    \\\hline
			$1000  $&$ 2     $&$ [100, 70000]    $&$ -     $&$ 23.570 $&$ 178   $& $(128,128,128)$     &$ 0.05             $                   \\
			$1000  $&$ 4     $&$ [100, 1000]     $&$ -     $&$ -     $&$ 145.8 $& $(192,192,128)$     &$ 0.05              $                   \\
			$1000  $&$ 8     $&$ [30, 10000]     $&$ -     $&$ 39.36  $&$ 143.1 $& $(256,256,192)$     &$ 1                  $                 \\\hline
			$10000 $&$ 2     $&$ [1000, 50000]   $&$ -     $&$ 156.577$&$ 495.1 $& $(192,192,384)$     &$ 0.05                $                
		\end{tabular}
		\caption{Table of simulation parameters and output data. The (horizontal) fit to the data for $\nu_E$ for very low frequencies is used to obtain $\alpha = \overline{\left(\frac{\nu_E}{u_z^{\text{rms}}d} \right)}$ (which can only be clearly obtained in cases with $R\leq100$). Also shown is the $y$-intercept of the fit to the data in the high frequency regime for which $\nu_E\propto(\omega/\omega_c)^{-2}$. We also report the volume-averaged RMS convective (vertical) velocity component, $u_z^{\text{rms}}=d\omega_c$. The table features various values of $R$ with a range of different domain sizes and reports the typical resolutions $(N_x,N_y,N_z)$ used for each set of simulations. Effects of the shear amplitude, $a_0$, are neglected in these results but we report the range of values explored.}
		\label{table_allsims}
	\end{table*}

	\begin{table}
		\begin{tabular}{ r | c | c | c }
			R 		& $a_0$	& $\gamma = \overline{\nu_E (\omega/\omega_c)^2}$  & $u_z^{\text{rms}}$ \\ \hline
			$100$ 	& $0.05$& $11.403$     & $48.83$    \\ \hline
			$100$ 	& $0.5$ & $29.282$     & $48.47$    \\ \hline
			$100$ 	& $1$ 	& $133.38$     & $49.93$    \\ \hline
		\end{tabular}
		\caption{Table listing values of $\gamma = \overline{\nu_E (\omega/\omega_c)^2}$, which is the $y$-intercept of the linear fit to the high frequency regime assuming a $(\omega/\omega_c)^{-2}$ power law, and the time and volume averaged vertical component of the velocity (the convective velocity), $u_z^\mathrm{rms}$, for various shear amplitudes, $a_0$. All cases are for $R=100$ and with domain size $(8,8,1)$. This shows the effects of varying the tidal amplitude on our high-frequency results for $\nu_E$.}
		\label{table_amp_depend}
	\end{table}

	In Table~\ref{table_allsims} we summarise three key quantities of interest from our simulations and the range of parameters explored. The key quantities are:
	\begin{itemize}
		\item time averaged rms of the vertical component of velocity, $u_z^{\text{rms}}$.
		\item linear fit values of $\alpha$	in the low frequency regime, indicating the constant of proportionality between $\nu_E$ and $u_z^{\text{rms}} d$.
		\item $\gamma = \overline{\nu_E(\omega/\omega_c)^2}$, which represents the $y$-axis crossing of the quadratic fit to the high frequency regime, for $R=100$, $a_0\in\{0.05,0.5,1\}$, including only simulations such that $\omega/\omega_c > 1$.
	\end{itemize}
	In the interests of examining the amplitude dependence of the effective viscosity in the high frequency regime, Table~\ref{table_amp_depend} lists $\gamma$ and $u_z^{\text{rms}}$ for the cases with $R=100$ in a domain of size $(8,8,1)$ for three different amplitudes $a_0\in\{0.05,0.5,1\}$. 
	
	In the larger domains, the low frequency regime is shifted to significantly lower frequencies than in the cases in \citetalias{duguid_tidal_2019} which makes this regime computationally difficult to examine. Where possible we report the linear fit to the low frequency regime to provide an estimate of $\alpha$ there. It is clear that as the domain size increases then the magnitude of the linear fit to $\alpha$ also increases until we reach an approximate convergence once the energetically dominant modes of the convection are contained in the box. This convergence can be seen most clearly in the cases with $R=100$ in Table~\ref{table_allsims}.
	
	In the high frequency regime we examine the quantity $\gamma$ and find that it increases with $R$ in chaotic and turbulent cases but appears to have a smaller value than in the laminar, deterministic cases ($R=2$ small domains $L_x\leq8$). In Table~\ref{table_amp_depend} we see that $\gamma$ also increases with increasing shear amplitude.
	
	As would be expected the data in Table~\ref{table_allsims} shows that as $R$ increases so does $u_z^{\text{rms}}$. For increasing domain size the values of $u_z^{\text{rms}}$ converge once $L_x\approx4$, which we note is similar to when the peak of the wavenumber spectrum is contained within the box. Table~\ref{table_amp_depend} shows that the shear amplitude has little, if any, effect on $u_z^{\text{rms}}$, which is used to scale the shear frequency. As such this, the amplitude dependence of $\gamma$ to be unlikely to be due to the shear significantly modifying the convection.

	\subsection{Spatial structure of the Reynolds stress and effective viscosity}
	
	To explore the mechanism governing the interaction between tides and convection in more detail, we analyse the spatial (wavenumber) spectrum of the Reynolds stress $\hat{R}_{xy}(n_x,n_y)$ and effective viscosity $\hat{\nu}_E(n_x,n_y)$. These quantities are vertically-integrated and time-averaged spatial spectra that are computed as a function of the horizontal integer wavenumbers $n_x$ and $n_y$, as described in section~\ref{section_QoI}. The computation of $\hat{\nu}_E(n_x,n_y)$ requires sufficiently good temporal resolution that the time integral (as in equation~\ref{maths_eff_visc}) is accurately computed.
	
\begin{figure}
		\includegraphics[width=\columnwidth]{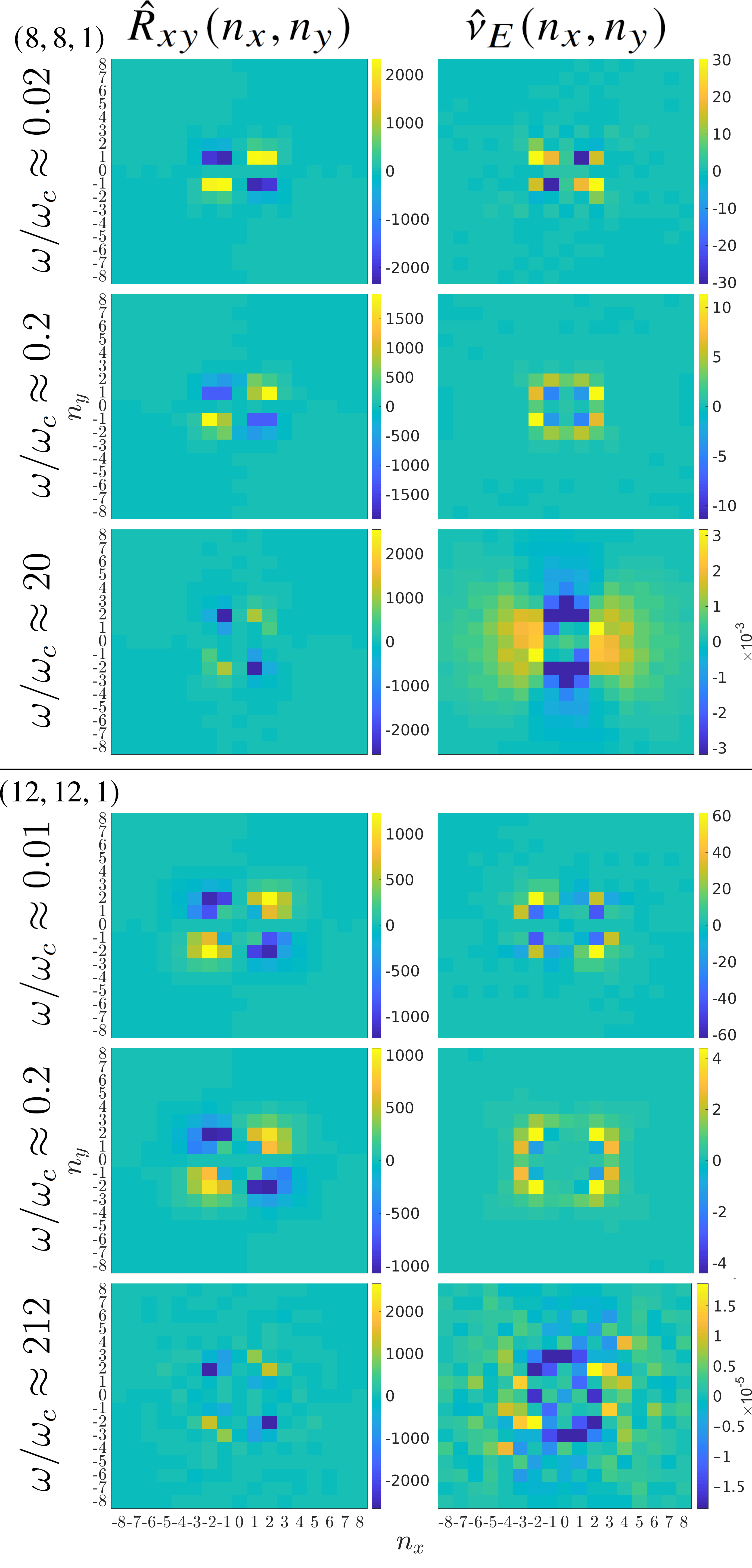}
		\caption{Examples of the temporally-averaged and vertically-integrated spatial spectra of the Reynolds Stress $\hat{R}_{xy}(n_x,n_y)$ (left column) and effective viscosity $\hat{\nu}_E(n_x,n_y)$ (right column) as a function of the integer wavenumbers $n_x$ and $n_y$. The top three rows are for $R=100$ with $\omega/\omega_c \in \{0.02, 0.2,20\}$ with a domain size of $(L_x, L_y, L_z) = (8,8,1)$. The bottom two rows are also for $R=100$ but with $\omega/\omega_c \in\{0.01,0.2,212 \}$, and for the larger domain size of $(L_x, L_y, L_z) = (12,12,1)$, which more clearly demonstrate the dominant ring in the spatial spectrum. The frequencies shown correspond to cases in the low, intermediate and high frequency regimes. Similar results are obtained for all values of $R$ and domain sizes explored.} 
		\label{figure_chessplots}
	\end{figure}

	Example cases are shown in figure~\ref{figure_chessplots} with $R=100$ and $a_0 = 0.5$ for two different domain sizes $(8,8,1)$ and $(12,12,1)$, each demonstrating three cases for each box size which lie in the low, intermediate and high frequency regimes (the respective frequencies can be seen in the figure). For adequate temporal averaging, we ensured that at least $10$ snapshots were taken per tidal period, and the simulations were integrated for many tens of tidal periods. Similar figures have been obtained for a number of other cases that show similar behaviour. We also note that the spectra in Fig.~\ref{figure_chessplots} have been zoomed in to show the lowest wavenumbers, since we find higher wavenumbers to contribute negligibly.
	
	We observe that $\hat{R}_{xy}(n_x,n_y)$ is maximal in a ring that coincides with the energetically-dominant wavenumber in Fig.~\ref{figure_boxes_resolution_spectra}, and this quantity then  falls off rapidly in magnitude with increasing $n_x$ and $n_y$. The same wavenumber ring also provides the dominant contribution to $\hat{\nu}_E(n_x,n_y)$. The modes in this ring provides the dominant contribution to the total effective viscosity $\nu_E$, suggesting that the largest (energetically-dominant) scales of the convection are the most important. This appears to contradict the main hypothesis of \cite{goldreich_turbulent_1977}, who claim that the resonant eddies dominate the interaction, and that the largest scales could at most contribute a comparable amount as the resonant eddies. There is a peak in the frequency spectrum (e.g. of the Reynolds stress) at the forcing frequency, but this does not appear to be correlated with a ring of modes in the wavenumber spectrum. Instead, it appears that it is the response of the energetically-dominant modes at the forcing frequency that dominates the contribution to $\nu_E$. However, we caution that our simulations do not possess a sufficiently long inertial range to clearly test the expectations of \cite{goldreich_turbulent_1977} solely within the turbulent cascade, which would require much more turbulent simulations.
	
	In the high frequency cases, the $\hat{\nu}_E(n_x,n_y)$ spectra shows a strong negative contribution from the nearly $x$-aligned components of the flow, and a slightly weaker contribution from the positive nearly $y$-aligned components. This is compatible with the predictions of the asymptotic theory in \citetalias{duguid_tidal_2019}. Note also that more modes provide an observable contribution to $\nu_E$ for larger frequencies, which results from the larger shear amplitude $a_0\omega$ in these cases.

	\section{Discussion}\label{section_discussion}
	
	In this paper we have presented a much wider parameter survey than \citetalias{duguid_tidal_2019}, and in particular we have studied convection in wider boxes, allowing the peak of the energy spectrum to be fully resolved. These new simulations support our prior results for high frequency tidal forcing, in that we find strong evidence in favour of $\nu_E\propto \omega^{-2}$, but they have also uncovered a new intermediate frequency scaling that has not been previously reported (or predicted). This new frequency scaling has $\nu_E\propto \omega^{-0.5}$ for frequencies $10^{-2}\lesssim \omega/\omega_c\lesssim 1-5$ (depending on amplitude). In this section we discuss further this new regime, and some of the implications of our results.
	
	Previous simulations of anelastic convection \citep{penev_direct_2009}, and Boussinesq spherical convection in a model with homogeneous internal heating \citep{vidal_turbulent_2020}, have observed an intermediate frequency scaling for $\nu_E\propto \omega^{-1}$ for a range of frequencies around $\omega\sim \omega_c$. This is consistent with the $-1$ slope in the frequency spectrum of the kinetic energy in the simulations of \cite{vidal_turbulent_2020} (and possibly also in \citealt{penev_direct_2009}). This differs from our results in this frequency range, where we find $\nu_E\propto\omega^{-0.5}$. In addition, simulations with various strengths of convective driving in spheres are found to give different exponents from -0.5 to -1 in the intermediate regime \citep{vidal_anomalous_2020}. Further work is required to explore in detail this difference, though we hypothesise that it may result from the radial variation in the heat flux in the spherical model, which is constant in our Cartesian case. 
	
	The study of the frequency spectrum of turbulent convection has been primarily directed towards the inertial range in order to make comparisons between the classical theories  \citep{kolmogorov_local_1941,bolgiano_turbulent_1959,obukhov1959effect}, which are based on the spatial spectrum, and experiments \citep{sano_turbulence_1989,ashkenazi_spectra_1999,wu_frequency_1990,shang_scaling_2001,liot_simultaneous_2016}, where the data is primarily temporal in nature, with the objective of understanding the nature of the turbulence. The low frequency portion of the spectrum has received far less attention, with the majority of prior interest coming from the classical area of ``$1/f$ noise" \citep{dmitruk_low-frequency_2007,pereira_1f_2019,vidal_anomalous_2020}. Our results suggest that an understanding of the frequency spectrum of convection may allow us to predict the effective viscosity acting on the equilibrium tide for low and intermediate frequencies (though perhaps not for high frequencies). As such, this provides new motivation for research into the long term dynamics of turbulent convection in more realistic models.

	The agreement of the frequency spectrum and the effective viscosity was observed to break down when the high frequency regime was reached. 	The transition to the high frequency regime depends on the tidal amplitude, where larger amplitudes are found to shift the transition to higher frequencies. This may be related to the relative energy in the tidal shear to the convection at these frequencies. However, this should be explored further in a future investigation. 
	
	Despite the existence of a $-2$ power law in the frequency spectrum of convective turbulence \cite{Landau1987Fluid,kumar_applicability_2018}, which the effective viscosity follows, the effective viscosity trend maintains this power law even when the frequency spectrum transitions into the dissipation range with a much faster fall-off. This demonstrates that the effective viscosity does not follow the spectrum at high frequencies (at least in our simulations), and the agreement in the power law may be coincidental. In the theoretical prediction of \cite{goldreich_turbulent_1977} the $-2$ power law was predicted by applying Kolomogorov turbulence and assuming that the ``resonant eddies" that are resonant with the tidal shear would would provide the dominant contributions to the effective viscosity. However, we have shown that a turbulent cascade is not required to obtain a $-2$ scaling (see also \citealt{ogilvie_interaction_2012,braviner_stellar_2015}). For example, 
	$R=2$ cases possess no inertial range in the wavenumber spectrum, which is hence non-Kolmogorov-like, and yet we still obtain a $-2$ power law for $\nu_E$. An independent prediction of the $-2$ scaling was made using asymptotic analysis \citep{ogilvie_interaction_2012} which we extended in our \citetalias{duguid_tidal_2019} to include thermal effects, which also allows for the prediction of negative effective viscosities.
	
	\cite{goldreich_turbulent_1977} claimed that the ``resonant modes" provide the dominant contributions to the effective viscosity, but the largest scale modes could contribute a comparable amount. We have conducted a Fourier analysis of the spatial structure of the Reynolds stress and of the contributions to the effective viscosity. We found that the effective viscosity is dominated by the energetically-dominant ring of modes in wavenumber space. We do not observe any appreciable contribution from resonant eddies. We do however observe a significant temporal resonance observed in the frequency spectrum (which is found to occur for all spatial wavenumber bins above the dissipation lengthscale).
	
	In this paper, and in \citetalias{duguid_tidal_2019}, we provided robust measurements of negative effective viscosities, as originally found in a slightly different convection model by \citet{ogilvie_interaction_2012}. Here we find that increasing the strength of the convection shifts the transition to higher frequencies, suggesting that for realistic Rayleigh numbers in planets and stars, the frequency required to produce a negative $\nu_E$, and therefore tidal anti-dissipation, would be prohibitively high (see also \citealt{vidal_turbulent_2020}). The negative values may therefore not be relevant in reality.
	
	\section{Astrophysical implications}
	\label{Implications}
	
	\begin{figure}
		\begin{center}
			\subfigure{\includegraphics[trim=3.5cm 0cm 5.5cm 1.03cm,clip=true,width=0.4\textwidth]{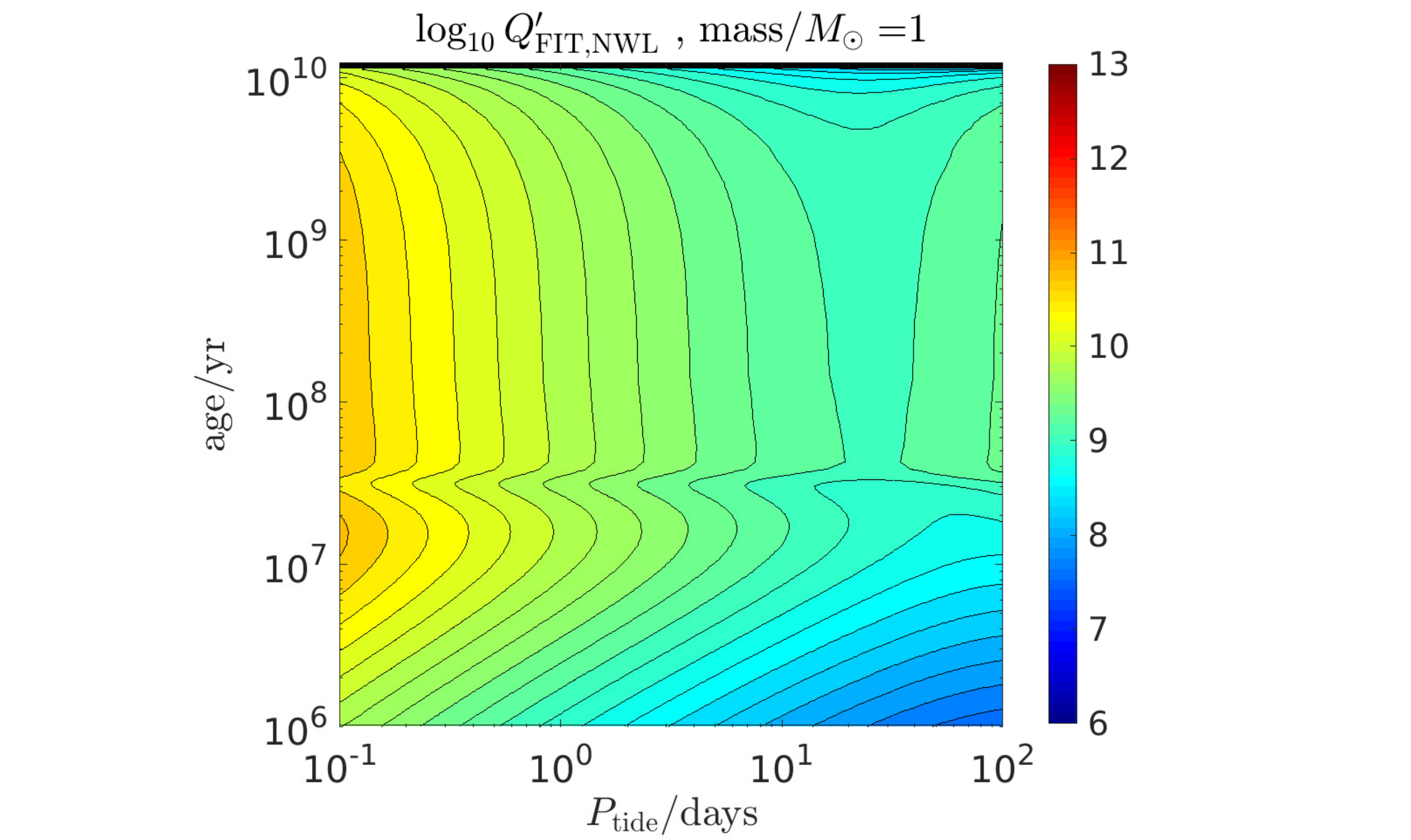}}
			\subfigure{\includegraphics[trim=1cm 0cm 3.5cm 0cm,clip=true,width=0.45\textwidth]{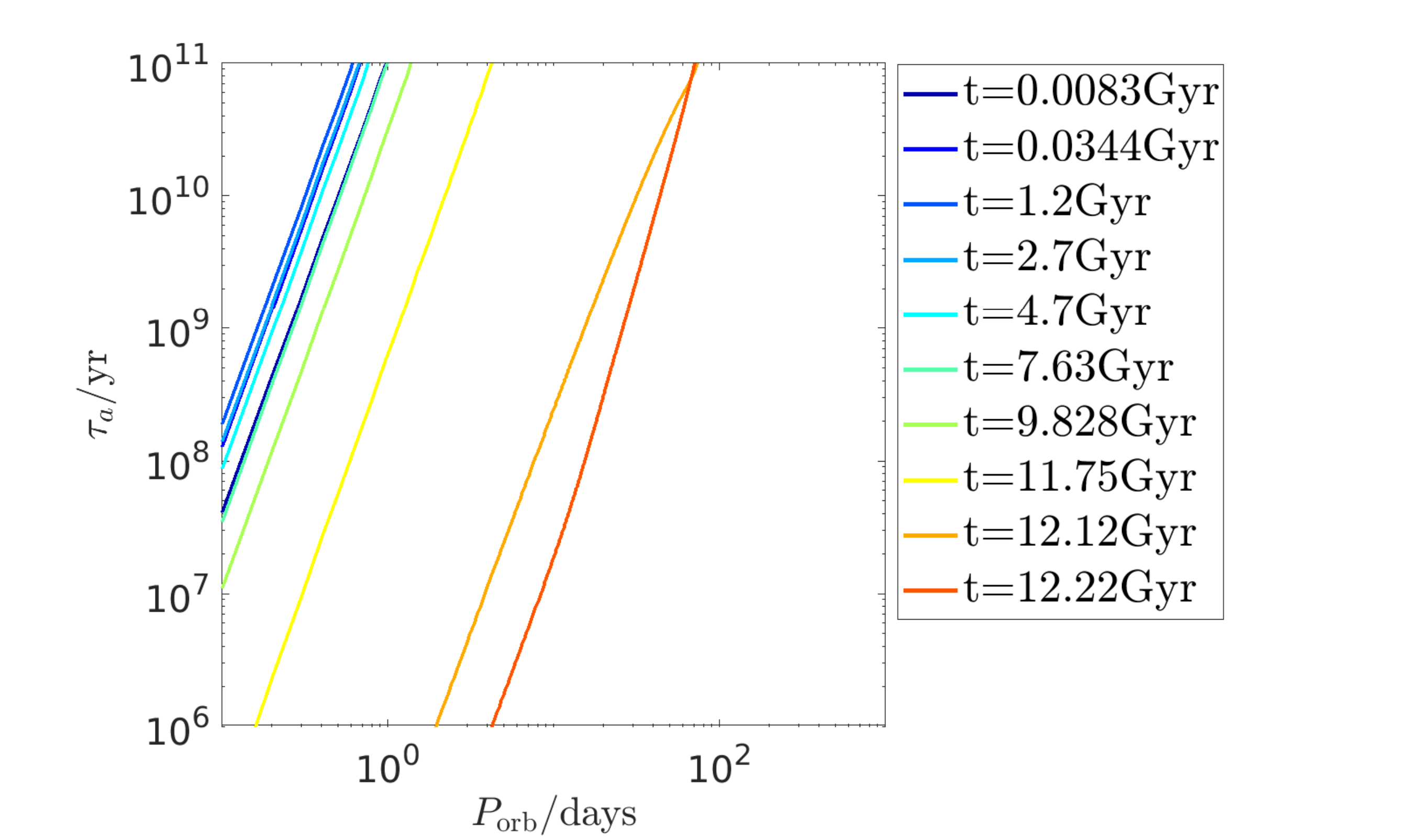}}
		\end{center}
		\caption{Top: Contours of $\log_{10} Q'_\mathrm{eq}$ resulting from dissipation of the correct equilibrium tide in the convective envelope of a $1 M_\odot$ star (with initial metallicity 0.02) as a function of age and tidal period based on applying $\nu_\mathrm{FIT}$. Bottom: inspiral time $\tau_a$ as a function of orbital period resulting from this mechanism, for a $1M_J$ planet in circular orbit about a slowly rotating $1M_\odot$ star at various ages prior to the red giant phase ($P_\mathrm{rot}=100$ d for all curves, which may be relevant for later ages).}
		\label{QpNWL}
	\end{figure}
	
	In many astrophysical applications, tidal forcing occurs in the high frequency regime for the dominant convection eddies, such that $\omega/\omega_c\gg 1$. For example, the tidal interaction of a hot Jupiter on a 1 d orbit around a slowly rotating solar-type star has a tidal period of $0.5$ d, but the convective eddies at the base of the convection zone have turnover timescales of order 20 d. Based on our results, this implies a significant reduction in the effective viscosity. As stars evolve, their convective velocities and length-scales evolve, leading to large changes in turbulent viscosities predicted by MLT. Here we apply our results to predict planetary orbital decay around an evolving solar-mass star.
	
	To apply our results, we must adopt an appropriate fit for $\nu_E$. We choose to fit the points in Fig.~\ref{figure_alpha_many_R} which provide the maximum estimate of the dissipation (here we ignore any possible amplitude dependence for $\omega/\omega_c\gtrsim 1$), such that we define
	\begin{eqnarray}
		\nu_{\mathrm{FIT}} =  u^{\mathrm{mlt}} l^{\mathrm{mlt}} \begin{cases}
			5 \quad & (\frac{|\omega|}{\omega_{c}}<10^{-2}), \\
			\frac{1}{2}\left(\frac{\omega_{c}}{|\omega|}\right)^{\frac{1}{2}} \quad &(\frac{|\omega|}{\omega_{c}} \in [10^{-2},5]), \\
			\frac{25}{\sqrt{20}}\left(\frac{\omega_{c}}{|\omega|}\right)^2 \quad &(\frac{|\omega|}{\omega_{c}}>5), \\
		\end{cases}
		\label{nuEFIT1}
	\end{eqnarray}
	$u^{\mathrm{mlt}}$ is the convective velocity and $l^{\mathrm{mlt}}$ is the mixing length, and $\omega_c=u^{\mathrm{mlt}}/l^{\mathrm{mlt}}$, which are obtained in stellar models computed with MESA \citep{Paxton2011, Paxton2013, Paxton2015, Paxton2018, Paxton2019}. For the purposes of this crude application (and in the absence of compressible simulations), we have simply replaced $d$ with $l^{\mathrm{mlt}}$ and $u_z^\mathrm{rms}$ with $u^{\mathrm{mlt}}$ and calculated these at each radius in the convection zone. To apply this in a stellar model we calculate the correct equilibrium tide in convective envelopes \citep{Terquem1998,ogilvie_tidal_2014}, and then compute the dissipation integral following the procedure outlined in\footnote{See also \cite{Zahn1989} and \cite{Remus2012} for a similar approach using the conventional equilibrium tide, which is strictly invalid in convection zones. This predicts more efficient dissipation than our model by a factor of 2-3 for the same $\nu_\mathrm{FIT}$.} \cite{B2020}. The result is then converted into a tidal quality factor $Q'_\mathrm{eq}$, which is an inverse measure of the dissipation \citep[e.g.][]{ogilvie_tidal_2014} (see \citealt{B2020} for further details).
	
	We show $Q'_\mathrm{eq}$ computed using $\nu_\mathrm{FIT}$ in models of a $1 M_\odot$ star (computed with an initial metallicity 0.02) in the top panel of Fig.~\ref{QpNWL} as function of tidal period $P_\mathrm{tide}=2\pi/\omega$ and age (in yrs). This shows that during the main sequence, solar-mass stars have $Q'_\mathrm{eq}\sim 10^{10}$ for tidal periods of order 1 day, though $Q'_\mathrm{eq}$ is smaller during pre-main sequence phases (ages prior to approximately $10^7$ yr) and as the star evolves off the main sequence (ages approaching $10^{10}$ yr), indicating more efficient dissipation during these phases.
	
	The resulting orbital decay rate for a (circularly orbiting) short-period planet of mass $M_p$ around a slowly rotating star of mass $M$ and radius $R$ can be computed from
	\begin{eqnarray}
		\label{dlnadt}
		\frac{\mathrm{d}\ln a}{\mathrm{d} t} 
		= -\frac{9\pi}{Q'_\mathrm{eq}}\left(\frac{M_p}{M}\right)\left(\frac{M}{M+M_p}\right)^{\frac{5}{3}}\frac{P_\mathrm{dyn}^{\frac{10}{3}}}{P_\mathrm{orb}^{\frac{13}{3}}},
	\end{eqnarray}
	where $a=\left(G(M+M_p)P_\mathrm{orb}^2/(4\pi^2)\right)^{1/3}$ is the semi-major axis, $P_\mathrm{orb}$ is the orbital period (assumed to be much shorter than the stellar spin period $P_\mathrm{rot}$), and $P_\mathrm{dyn}=2\pi \sqrt{R^3/GM}$ is the dynamical timescale. We note that the relevant tidal frequency, assuming $P_\mathrm{orb}\ll P_\mathrm{rot}$ is $\omega=4\pi/P_\mathrm{orb}$, and that the high-frequency regime is found to be the relevant one for short orbital periods (at all stages in stellar evolution). As a result $\nu_\mathrm{FIT}\propto P_\mathrm{orb}^2$, implying that $Q'_\mathrm{eq}\propto P_\mathrm{orb}^{-1}$, so that the right hand side of Eq.~\ref{dlnadt} is proportional to $a^{-5}$, indicating accelerating inspiral. The corresponding timescale for orbital decay of a 1 $M_J$ hot Jupiter is 
	\begin{eqnarray}
		\label{taua}
		\tau_a \approx -\frac{1}{5} \left(\frac{\mathrm{d}\ln a}{\mathrm{d} t}\right)^{-1} \approx 250 \,\mathrm{Gyr} \left(\frac{M_p}{M_J}\right) \left(\frac{P_\mathrm{orb}}{1\,\mathrm{d}}\right)^{\frac{10}{3}},
	\end{eqnarray}
	which we have evaluated in a stellar model similar to the current Sun (assuming $P_\mathrm{rot}=20$ d). 
	
	We show similar estimates for $\tau_a$ as a function of orbital period $P_\mathrm{orb}$, computed numerically for a $1M_\odot$ star for a range of ages, in the bottom panel of Fig.~\ref{QpNWL}. We have assumed $P_\mathrm{rot}=100$ d for all ages for the purposes of this figure, since the curves shown are unaffected by rotation except for the latest ages when the star is expected to rotate so slowly. This figure, and the estimate in Eq.~\ref{taua}, indicates that convective damping of equilibrium tides plays a negligible role for planetary orbital decay around main sequence stars, even assuming the most optimistic fit for $\nu_\mathrm{FIT}$ consistent with our simulations. This is because of the strong reduction in the effective viscosity with the quadratic scaling law. On the other hand, for later evolutionary stages as the star begins to evolve onto the red giant phase, this mechanism becomes more efficient, primarily because the stellar radius becomes much larger. This mechanism thus predicts the destruction of many short-period planets during the later stages in the evolution of solar-mass stars. (We have omitted figures showing even later evolutionary stages for clarity, but planets out to much wider orbits can be rapidly destroyed by this mechanism.)

	Our results using $\nu_\mathrm{FIT}$ (with $\alpha=5$ at low frequencies) is found to predict more efficient dissipation by approximately a factor of 15, and therefore shorter tidal evolutionary timescales by this factor, than the usual assumption $\alpha=1/3$ that is usually assumed when applying the  \cite{goldreich_turbulent_1977} reduction \citep[e.g.][]{OgilvieLin2007}. We note that an enhancement in $\alpha$ is apparently required to explain the results of \cite{Hansen2012}, though they employ the linear reduction law, and thus their model predicts much more efficient dissipation at high frequencies. However, we caution that the application here of our results in Fig.~\ref{figure_alpha_many_R}, based on Boussinesq simulations, is very crude. In addition, our simulations find a scatter at high frequencies of approximately an order of magnitude depending on tidal frequency, so the precise results of our application are probably uncertain to within at least such a factor.

	\section{Conclusions}\label{section_conclusions}
	
	The interaction between large-scale equilibrium (non-wavelike) tidal flows and turbulent convection is thought to be an important mechanism of tidal dissipation in giant planets and stars. However, it is probably the most uncertain tidal mechanism, making it difficult to make robust predictions for the resulting spin-orbit evolution in astrophysical systems. In particular, it is thought that the effective viscosity mediating the interaction between the tidal flow and convection depends strongly on the tidal frequency, and its efficiency is expected to be greatly reduced when the tidal frequency is larger than the relevant convective frequency \citep{zahn_les_1966,goldreich_turbulent_1977,GoodmanOh1997}. However, the correct frequency scaling that should be applied in the high frequency regime has been a matter of much controversy, with the original work of \cite{zahn_les_1966} proposing $\nu_E \sim \omega^{-1}$ when $\omega\gg \omega_c$, and \cite{goldreich_turbulent_1977} later proposing $\nu_E \sim \omega^{-2}$ instead. It is essential to resolve this issue, and to determine the correct frequency-dependence of the effective viscosity, before we can apply this mechanism to make robust predictions for tidal evolution in planetary systems and binary stars.

	We have presented the results from an extensive parameter survey of numerical simulations designed to explore the interaction between large-scale equilibrium tidal flows and convection within a star or giant planet. We have used Boussinesq hydrodynamical simulations of a local Cartesian patch of convective fluid, which is modelled within the well-studied Rayleigh-B\'{e}nard system, to which we impose a large-scale tidal-like shear flow as a ``background flow". Our analysis of these simulations has primarily focused on the evaluation of the effective viscosity which arises as a result of the interaction between the convection and this tidal-like flow. We have presented an in-depth study into the relationship between the frequency spectrum of both the	energy and the Reynolds stress in the convection and the frequency-dependence of the effective viscosity. This parameter survey is a direct extension of \citetalias{duguid_tidal_2019}, and is guided by and builds upon the results therein, as well as those of \cite{penev_direct_2009,ogilvie_interaction_2012} and \cite{vidal_turbulent_2020}. In particular, we have explored a wider range of parameters than \citetalias{duguid_tidal_2019}, to explore the dependence of the effective viscosity on tidal frequency and amplitude, as well as the Rayleigh number and domain size.
	
	We have determined that the effective viscosity governing the interaction between tidal flows and convection exhibits three different regimes depending on the ratio of the tidal and convective frequencies (as shown in e.g.~Fig.~\ref{figure_alpha_many_R}). We refer to these as the low frequency, intermediate frequency and high frequency regimes. Our main results are as follows, where we also highlight which of the three regimes each statement applies to:	
	\begin{itemize}
		\item[1.] (\emph{low frequency regime}) For very low tidal frequencies, the effective viscosity becomes frequency-independent. The transition into this regime occurs at $\omega/\omega_c \lesssim 10^{-2}$, which is a much lower frequency than has been predicted \citep{zahn_les_1966,goldreich_turbulent_1977} or observed in simulations to date (\citetalias{duguid_tidal_2019}). Previous work instead expected or observed the transition to occur at approximately the convective frequency. This frequency-independent regime coincides with the commonly-adopted constant tidal time-lag model \citep[e.g.][]{Darwin1880,Mignard1980,Hut1981,Eggleton1998}, which our results have shown is only valid for a limited range of very low tidal frequencies $\omega/\omega_c \lesssim 10^{-2}$. The constant time-lag model is therefore not appropriate for modelling tidal interactions except for such low frequencies, which are usually not relevant in astrophysical applications.
		
		\item[2.] (\emph{low frequency regime}) We find this mechanism to be considerably more efficient than has been previously proposed at very low frequencies. In particular, we have determined that $\nu_E\approx \alpha u_z^{\text{rms}} d$, where\footnote{We remind the reader that this is strictly different from the usual mixing length ``$\alpha$'' parameter, see footnote \footref{footnote_alpha}.} $\alpha\approx 5$. This result appears to be independent of Rayleigh number, suggesting that we might be able to extrapolate this to astrophysical parameter values. Previous work has adopted a naive mixing-length picture based on the analogy with kinetic theory, which instead gives $\alpha=1/3$ \citep[e.g][]{Zahn1989,OgilvieLin2007} if $d$ corresponds with the usual mixing length. In our Boussinesq model $d$ is the most natural length scale to identify with the mixing length, but compressible models are needed before we can be fully confident of the appropriate value of $\alpha$.
		
		\item[3.] (\emph{intermediate regime}) We have discovered a new regime with a different frequency scaling $\nu_E\propto \omega^{-0.5}$, which occurs in the range $10^{-2} \lesssim \omega/\omega_c \lesssim 1-5$ (depending on tidal amplitude), which we refer to as the intermediate-frequency regime. This regime is observed for all Rayleigh numbers considered, suggesting that it might be a robust feature of the interaction between tides and convection. To the best of our knowledge, this regime has never previously been predicted or reported. A similar intermediate regime, but with a different power law of $-1$ was however observed for spherical convection in \cite{vidal_turbulent_2020} for $1 \lesssim \omega/\omega_c \lesssim 5$. The existence of such an intermediate regime here and in \cite{vidal_turbulent_2020} may explain the previous disagreement between \citetalias{duguid_tidal_2019} and \cite{ogilvie_interaction_2012} compared with \cite{penev_direct_2009}. 
		This new regime may be relevant in many astrophysical applications where the constant time-lag model was previously applied.
		
		\item[4.] (\emph{low/intermediate regime}) The frequency scaling of the effective viscosity, in both the low and intermediate frequency regimes, appears to follow the corresponding slope of the
		frequency spectrum of the kinetic energy (and also the Reynolds stress) when $\omega/\omega_c \lesssim 1$ \citep[see also][]{vidal_turbulent_2020}. This is shown in Fig.~\ref{figure_nu_with_spectra}. In these regimes, the agreement of the slope of the eddy viscosity points with both the energy and Reynolds stress curves is robust, but the constants of proportionality could depend on the Rayleigh number, the Prandtl number and the tidal amplitude (though the dependence on the latter has been found to be weak). In principle, this could be determined by performing a more extensive parameter survey for larger $R$ for frequencies that lie within the intermediate regime. This would be a challenging task however, since simulations with large $R$ are computationally costly, and we have only been able to robustly find the intermediate regime for $\nu_E$ in simulations with $R \leq 500$.
		
		\item[5.] (\emph{high frequency regime}) For $\omega\gtrsim \omega_c$, we provide strong evidence clearly demonstrating that the effective viscosity follows $\nu_E\propto \omega^{-2}$, in agreement with prior simulations \citep{ogilvie_interaction_2012,braviner_stellar_2015,duguid_tidal_2019,vidal_turbulent_2020} and theoretical expectations \citep{goldreich_turbulent_1977,goldman_effective_2008,ogilvie_interaction_2012,duguid_tidal_2019}. This mechanism is therefore much less efficient for high frequency tidal forcing than would be predicted by adopting the less drastic frequency-reduction of  \cite{zahn_les_1966}. One implication is that this mechanism is unlikely to cause appreciable orbital decay for hot Jupiters orbiting main-sequence stars (for which dynamical tide mechanisms such as internal gravity wave damping are probably much more important).
		
		\item[6.] (\emph{high frequency regime}) Despite our simulations being in agreement with \cite{goldreich_turbulent_1977} in finding $\nu_E\propto \omega^{-2}$ in the high frequency regime, our results do not support their physical explanation. This is most clearly evident from our observation that it is the energetically-dominant modes of the convection which contribute the most to the effective viscosity. In fact, we do not observe any significant contribution from the ``resonant modes" in the spatial spectrum (see Fig.~\ref{figure_chessplots}), which were predicted by \cite{goldreich_turbulent_1977} to provide the dominant contribution. However, we do observe resonant behaviour in the frequency spectrum of the kinetic energy, particularly when the shear is in the high frequency regime. In the absence of a simple mechanism to explain this scaling, the asymptotic analysis in \cite{ogilvie_interaction_2012}, which we extended in \citetalias{duguid_tidal_2019}, does however provide a mathematical prediction for this behaviour.
	\end{itemize}

	Despite much progress having being made in recent years on this problem there is still much work to do to understand the physics of the interaction between tidal flows and convection. \citetalias{duguid_tidal_2019} and this paper performed Boussinesq simulations, which effectively limits them to small domains relative to a pressure scale height, but convection in stars can occur over many scale heights so that compressible effects could be important. We propose that simulations to investigate anelastic convection, which would build upon \cite{penev_direct_2009} by exploring a much wider range of parameters, and in particular tidal frequencies, would be of great interest. These simulations would be able to make a more quantitative comparison with mixing-length theory. In addition, since all stars and planets rotate, it is important to study the effects of convection in this problem. It is known that sufficiently rapid rotation acts to constrain convection \citep{Stevenson1979,BDL14,Currie2020}, which probably affects the effective viscosity \citep{mathis_impact_2016}. The consequences of the inclusion of rotation, and its effects on the frequency spectrum of kinetic energy, have not yet been explored numerically.

	\section*{Acknowledgements}
	We would like to thank the reviewer for a prompt and careful reading of the manuscript and for their helpful suggestions. CDD was supported by EPSRC CDT in Fluid Dynamics EP/L01615X/1. AJB was supported by STFC grants ST/R00059X/1 and ST/S000275/1. CAJ was supported by STFC grant ST/S00047X/1.
	
	This work was undertaken on ARC1, ARC2, ARC3 and ARC4, part of the High Performance Computing facilities at the University of Leeds, UK. Some simulations were also  performed using the UKMHD1 allocation on the DiRAC Data Intensive service at Leicester, operated by the University of Leicester IT Services, which forms part of the STFC DiRAC HPC Facility (www.dirac.ac.uk). The equipment was funded by BEIS capital funding via STFC capital grants ST/K000373/1 and ST/R002363/1 and STFC DiRAC Operations grant ST/R001014/1. DiRAC is part of the National e-Infrastructure.
	
	\section*{Data availability}
	The data underlying this article will be shared on reasonable request to the corresponding author.

	\bibliographystyle{mnras}
	\bibliography{mybib}

	\appendix
	
	\section{Additional material}
	\label{appendixB}
	\begin{figure*}
		\includegraphics[width=2\columnwidth]{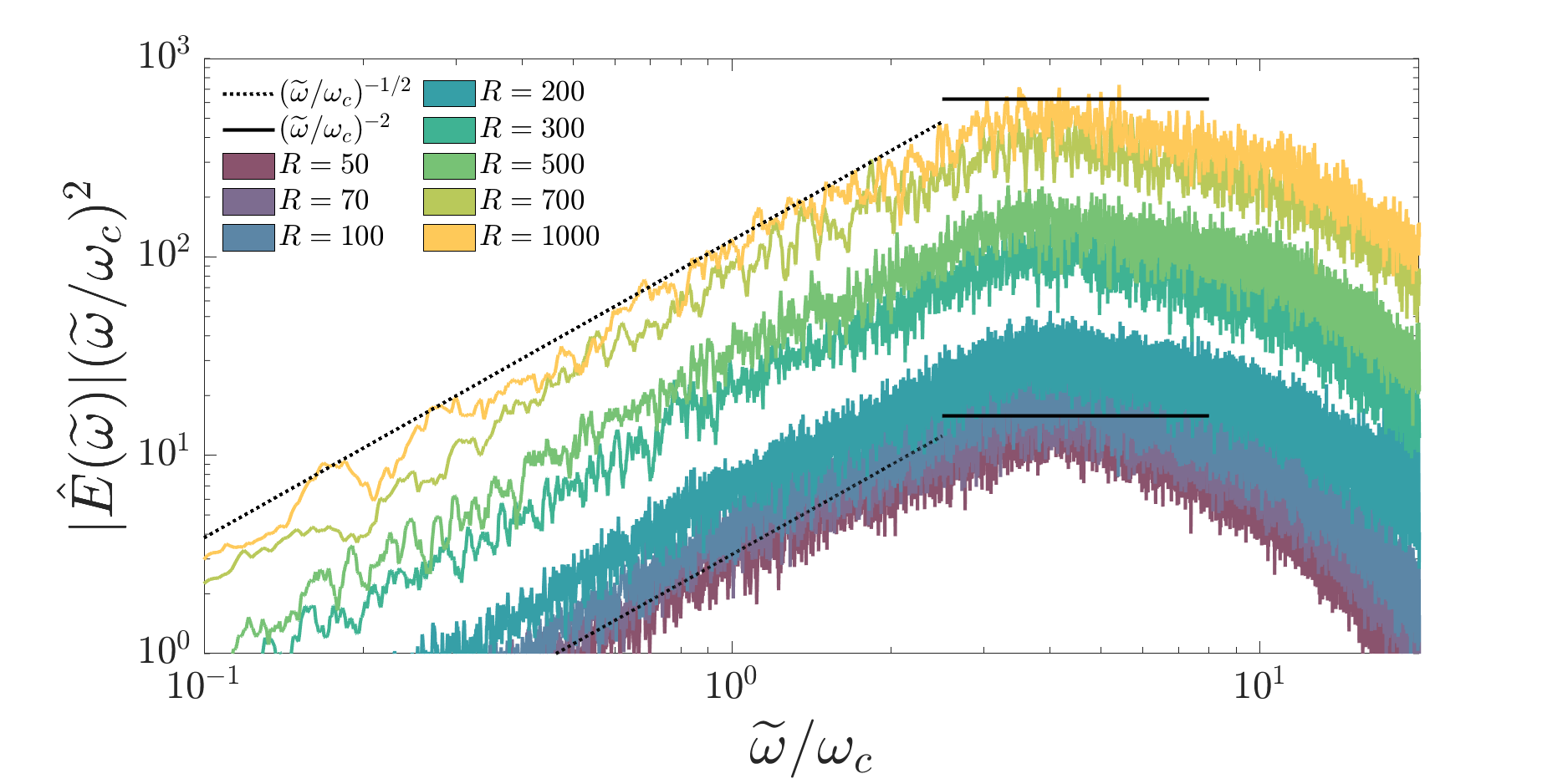}
		\caption{The same as Fig.~\ref{figure_kinetic_energy_noshear} but with the spectra scaled by $(\tilde{\omega}/\omega_c)^{2}$ in order to highlight the short inertial range. The inertial range for the $R=1000$ case is approximately $(3,6)$ while it is vanishingly small for $R=50$.}
		\label{figure_kinetic_energy_noshear_scaling}
	\end{figure*}

	\label{appendixA}
	\begin{table*}
	\begin{tabular}{c|c|c|c|c|c|c|c}
		$R$    & $L_x(=L_y)$ & $u_x^{\text{rms}}$  & $u_y^{\text{rms}}$   & $u_z^{\text{rms}}$   & $N_x=N_y$  & $N_z$  & $E_{\perp}$       \\\hline\hline
		2    & 2  & 5.44   & 0      & 5.44   & 32  & 32  & 29.6           \\
		2    & 4  & 3.64   & 7.2    & 4.29   & 64  & 32  & 44.4           \\
		2    & 8  & 8.01   & 3.43   & 4.64   & 64  & 32  & 52.6       \\
		2    & 12 & 5.19   & 5.59   & 4.11   & 128 & 64  & 38.7        \\
		2    & 16 & 5.55   & 5.33   & 4.07   & 256 & 64  & 38.3        \\
		2    & 24 & 5.47   & 5.5    & 4.05   & 512 & 64  & 38.6        \\
		2    & 32 & 5.54   & 5.45   & 4.05   & 512 & 64  & 39.1        \\\hline
		100  & 2  & 47.67  & 51.11  & 64.32  & 64  & 128 & 4637       \\
		100  & 4  & 61.01  & 59.81  & 48.98  & 64  & 128 & 4919      \\
		100  & 8  & 56.99  & 58.37  & 48.61  & 128 & 128 & 4527.4     \\
		100  & 12 & 58.49  & 58.03  & 48.11  & 192 & 128 & 4589.9     \\
		100  & 16 & 58.18  & 58.49  & 47.52  & 256 & 128 & 4570.2    \\
		100  & 24 & 58.29  & 58.8   & 47.71  & 384 & 128 & 4591      \\
		100  & 32 & 58.63  & 58.68  & 47.71  & 512 & 128 & 4604.4   \\\hline
		1000 & 2  & 148.94 & 146.66 & 178.74 & 128 & 192 & 38783       \\
		1000 & 4  & 166.22 & 173.82 & 145.43 & 192 & 192 & 39643       \\
		1000 & 8  & 164.96 & 164.22 & 144.64 & 256 & 192 & 37302      \\
		1000 & 12 & 165.81 & 167.59 & 143.7  & 384 & 192 & 37937      \\
		1000 & 16 & 167.66 & 165.68 & 143.29 & 512 & 192 & 37963     
	\end{tabular}
	\caption{Table listing the time-averaged RMS velocity components $u_i^{\text{rms}} : i \in \{x,y,z\}$, and the horizontal $N_x=N_y$ and vertical $N_z$ resolutions, for each $R$ and domain size. We evaluate the energy per unit area $E_{\perp}$ in each case. This table is associated with the un-sheared cases of convection reported in Fig.~\ref{figure_kinetic_energy_noshear}.}
	\label{table_allsims_noshear}
\end{table*}

	\bsp	
	\label{lastpage}
\end{document}